\documentclass[preprint,10pt,authoryear]{elsarticle}
\usepackage[a4paper, left=1cm, right=1cm, top=1cm, bottom=1.5cm]{geometry} 
\usepackage{amsmath}
\usepackage{graphicx}
\usepackage{caption}
\usepackage{epstopdf}
\usepackage{flushend}
\usepackage{color}
\usepackage[backref]{hyperref}
\usepackage{setspace}
\usepackage{algorithmicx,algorithm}
\usepackage{algpseudocode}
\usepackage{subfigure}
\usepackage{booktabs,multirow,longtable}
\usepackage{amssymb}
\usepackage{multicol}

\newcommand{\tabcaption}{\def\@captype{table}\caption}
\newcommand{\figcaption}{\def\@captype{figure}\caption}
\usepackage{natbib}
\usepackage{multicol} 
\floatstyle{plain}
\newfloat{myalgo}{tbhp}{mya}

\makeatletter
\newenvironment{tablehere}
    {\def\@captype{table}}
    {}
\newenvironment{figurehere}
    {\def\@captype{figure}}
    {}
\makeatother

\floatstyle{plain}

\journal{Arxiv}

\begin{document}

\begin{frontmatter}


\title{Community Detection of Complex Network Based on Graph Convolution Iterative Algorithm}

\author{Jiaqi Yao$^{a,b*}$ and Lewis Mitchell$^{a}$}
\address[label1]{School of Computer and Mathematical Sciences, the University of Adelaide 5005, Australia.} 

\address[label2]{Haide College, Ocean University of China, Qingdao 266071, China}
\cortext[cor1]{Corresponding author. \\ \text{~~~~}E-mail address: jiaqi.yao@student.adelaide.edu.au (J. Yao), lewis.mitchell@adelaide.edu.au (L. Mitchell).}

\begin{abstract}
Community detection can reveal the underlying structure and patterns of complex networks, identify sets of nodes with specific functions or similar characteristics, and study the evolution process and development trends of networks. Despite the myriad community detection methods that have been proposed, researchers continue to strive for ways to enhance the accuracy and efficiency of these methods. Graph convolutional neural networks can continuously aggregate the features of multiple neighboring nodes and have become an important tool in many fields. In view of this, this paper proposes a community detection method for complex networks based on graph convolution iteration algorithm. Firstly, the candidate community centers are determined by random sampling and the node attribute matrix is obtained based on the distances of nodes to community centers. Next, the graph convolution operation is implemented to obtain the convolutional node attribute matrix. Then, community partitioning method according to the convolutional node attribute matrix is presented and the effectiveness of community partitioning is measured through modularity. The method proposed in this paper is applied into to multiple random and real-world networks. The comparison results with some baseline methods demonstrate its effectiveness.
\end{abstract}


\end{frontmatter}



\begin{multicols}{2}

\section{Introduction}

Many complex systems in practice can be modeled as complex networks, such as social networks \citep{2022Influence}, biological networks\citep{2022Exploring} and transportation networks \citep{2019Feng}. The research on complex networks has received widespread attention, especially with the flourishing development of various online social platforms. Complex networks frequently display discernible structural characteristics, such as small-world and scale-free properties, along with community structures \citep{2015Complex}. 

The purpose of community detection is to identify densely connected subgroups or communities within the network \citep{2016Community}. By decomposing the network into different communities, it is possible to reveal the underlying structure and patterns of the networks, identify sets of nodes with specific functions or similar characteristics, and study the evolution process and development trends of networks. Community detection plays a critical role in various domains, including criminal gang identification \citep{2014Detecting}, product recommendation \citep{2016Context}, and stock price prediction \citep{2023MuDCoD}. Therefore, it has important theoretical and practical significance. 

Although the issue of community detection in complex networks has been thoroughly investigated, leading to the development of many standard algorithms \citep{2002Community}\citep{2018Community}, the pursuit of methods that are both more efficient and accurate remains an important topic in this field .

Graph Convolutional Neural Networks (GCNN) is a deep learning model used for processing graph structured data, which learns node attributes by performing convolution operations on the graph, taking into account the relationships between nodes. In recent years, GNNs have been successfully applied to problems such as node classification, graph classification and link prediction \citep{2023Graph}Graph\citep{2021Graph}.

However, conventional GCNNs typically entail training on weight matrices. Since each network varies in terms of the numbers of nodes and edges, and structure, it is necessary to train personalized weight matrices for each network when applying GCNN in community detection problems. Therefore, there may be significant costs involved.

In view of this, this paper proposes a community detection method of complex networks based on graph convolution iteration algorithm, which retains the advantages of conventional GCNNs and eliminates the computational complexity required for training weights. In addition, we have also made corrections to the convolution operation. This algorithm mainly consists of five parts: initialization of node attribute matrix, graph convolution operation, community partitioning method, evaluation of partitioning results, and algorithm termination conditions. 

This paper has the following three-fold innovations:

\begin{enumerate}
  \item  A method to measure node proximity using the distance between nodes is proposed, and based on which the node attribute matrix is constructed.

  \item  A community detection method based on graph convolution iterative algorithm is presented. The GCNN used in this paper has essential differences from traditional ones. Firstly, we do not convert the diagonal elements of the adjacency matrix from 0 to 1. Secondly, our algorithm does not need to train the weight matrix. Finally, we mainly rely on controlling the number of convolutions to obtain the optimal community partitioning results, rather than setting the number of convolutions beforehand.

  \item  The proposed method is applied to different random and real-world networks, and comparison with some baseline methods is conducted to prove its effectiveness.
  
\end{enumerate}

The remainder of this paper is organized as follows: The related work is overviewed in Section 2; Section 3 gives a preliminary; The community detection method base on graph convolution iteration algorithm is elaborated in Section 4. Section 5 provides a detailed introduction to the implementation process of the proposed algorithm through a case study. The experiment to validate the proposed method are put forward in Section 6. Finally, Section 7 summarizes the work of this paper.

\section{Related Work}

In 2002, Girven and Newman \citep{2002Community} first proposed the concept of community, which refers to densely connected subgroups in a network. So far, various community detection methods have been proposed, mainly divided into four categories: graph partitioning method, similarity based method, label propagation algorithm, and deep learning based method. 

\subsection{Graph Partitioning Method}

Graph partitioning methods divide the nodes of a network into groups and evaluate the quality of the partitioning result through an objective function. The larger the value of the objective function, the better the partitioning result is considered. The most commonly used objective function is modularity. 

KL algorithm \citep{1970An} is a efficient heuristic procedure to partition a network into two communities. GN algorithm \citep{2004Finding} uses the betweenness of edges as the similarity, and then removes edges with high betweenness each time. FN algorithm is an improvement of the GN algorithn \citep{M2004Fast}, which is a greedy fast community discovery method. Although FN has higher efficiency, its modularity decreases compared with GN. Blondel et al. \citep{2008Fast} proposed the Louvain algorithm based on hierarchical clustering, which is divided into two stages. Belim and Larionov \citep{2013Belim} introduced a greedy algorithm for community identification. 

LFVC \citep{Sabir2017} is a novel community partitioning algorithm that maximizes a new centrality metric, local Fiedler vector centrality. Zhang et al. \citep{Informationzhang} presented a large-scale community detection approach based on core node identification and layer-by-layer label propagation. Boroujeni and Soleimani \citep{2022Boroujeni} tried to identify influential nodes, and then the communities were detected by estimating their influence domain. Zhu et al. \citep{2023Multi} gave a multi-level community generation framework aimed at reducing the time complexity of community detection models in attribute networks. 

Due to its robust global search capability, the heuristic algorithms are extensively applied in modularity optimization within various fields \citep{2016Asurvey}. The algorithms involved include genetic algorithm \citep{2022Parallel}, evolutionary algorithm \citep{2023Reihanian}, particle swarm algorithm \citep{2021Label}, simulated annealing algorithm \citep{2016He}, etc. PODCD \citep{2021PODCD} is a probabilistic overlapping community detection method called  that used the block coordinate decent method to solve the objective function of the matrix factorization model. 

\subsection{Similarity Based Method}

Community detection methods based on similarity usually first evaluate the similarity between each pair of nodes, and then classify nodes with high similarity into the same community. The specific classification methods can be further divided into spectral clustering and density based methods.

The spectral clustering method is first proposed by Fiedler \citep{19731Fiedler}, which used the properties of the graph Laplacian matrix to segment the graph. This method requires manually setting the number and size of communities. 

The density based method determines whether nodes belong to the same community by their neighborhood density. The outstanding representative in this type of algorithm is DBSCAN algorithm \citep{1996Ester}. Compared with the classic clustering algorithm, K-means, DBSCAN has the advantage that it does not require a preset number of clusters and has no assumptions about the shape of the clusters. 

Matthieu and Pascal \citep{2005Computing} proposed a measure of similarities between vertices based on random walks and used it in an agglomerative algorithm to detect the community structure of a network. 
Lv et al. \citep{Lv0Detection} put forward a community detection algorithm based on proximity ranking and signal transmission, which calculates the similarity between nodes based on the idea of signal transmission. Hou et al. \citep{Hou2021} used the topological and semantic similarity between developers to construct a similarity matrix in the developer collaboration network.
 
\subsection{Label Propagation Algorithm}

Label propagation algorithm identifies communities by spreading labels or information across the network. This kind of methods are usually suitable for situations where there is a clear community structure in the network.

Raghavan et al. first applied label propagation algorithm for community structure information detection \citep{2007Near}. In order to overcome the shortcomings of traditional label propagation algorithms, an overlapping community detection method based on DeepWalk and improved label propagation \citep{9729518Yu} was proposed. Lu et al. \citep{8443129} put forward an improved label propagation algorithm based on the influence of adjacent nodes for identifying overlapping communities. Yazdanpara et al. \citep{Yazdanparast} presented a community detection acceleration modular gain method based on label propagation. A novel LPA algorithm \citep{9582769} was proposed to integrateLFVC multi-layer neighborhood overlap and historical label similarity. Zhao et al. \citep{2021Acommunity} presented a large-scale community detection method based on graph compression and label propagation algorithms. Gregory \citep{2009Finding} extended the label and propagation step to include information about more than one community.

\subsection{Deep Learning Based Method}

Based on deep learning methods, GCNNs and other deep learning models are used to learn the representation of nodes in the network, and based on this, community partitioning is carried out. 

Zhang et al. \citep{2019Attributed} proposed an adaptive graph convolution method for graph clustering that exploits high-order graph convolution to capture global cluster structure and adaptively selects the appropriate order for different graphs. Wang et al. \citep{2023Aknowledge} established a knowledge graph-GCN-community detection integrated model for large-scale stock price prediction, in which K-means, community detection and GCNs are merged to obtain accurate clustering results for similar stocks. Wu \citep{WuStudy} proposed an unsupervised community detection algorithm based on graph convolutional network model. The algorithm first selects some nodes in the graph to add artificial labels to simulate the input signal, and then passes the labels to adjacent nodes through the modified propagation rules of the graph convolutional network. Mohammed et al proposed a parallel deep learning-based community detection method in large complex networks (CNs), in which particle swarm optimization is used to improve the backpropagation \citep{NASSERALANDOLI202294}. 

Although some preliminary attempts have been made, the application of GCNNs in community detection problems is still in its early stages and requires extensive research.

The above research results have provided various ideas and methods for the community detection problem of complex networks, which greatly enriched the theory of complex networks and effectively improved the accuracy of community detection. However, the search for more efficient and accurate community detection methods is still the focus of attention in the field of complex network research at home and abroad, and the in-depth research in this popular direction will continue to be promoted on the basis of the existing research results.

\section{Preliminary} 

\subsection{Problem Description}

A complex network often serves as a simplified representation of a complex system, depicting individuals as nodes and capturing their relationships through edges, which are defined by specific rules or natural connections within the system. Generally, a network can be represented as a tuple, denoted as $G = (V, E)$, where $V = \{ {v_1},{v_2}, \cdots ,{v_n}\} $ is the node set, and $n$ representing the number of nodes; $E$ is the edge set, $(v_i, v_j) \in E$ indicates that there is an edge between $v_i$ and $v_j$, and $m=|E|$ represents the number of edges.

If there is an edge between $v_i$ and $v_j$, we say that they are adjacent, and one is called a adjacent of the other one. Let

\begin{equation}a_{ij}=\left\{\begin{array}{ll}
             1,& \text{if } v_i \text{ and } v_j \text{ are adjacent}\\
             0,& \text{otherwise} 
           \end{array}\right.
\end{equation}

\noindent then $A=(a_{ij})$ is called the adjacent matrix of the network. $d(v_i)=\sum_{j=1}^{n}a_{ij}$ is called the degree of $v_i$.

A community partition of $G$ is denoted as $\mathcal{C} = \{ C_i|C_i \in {2^{{V}}},i = 1, \cdots, \nu\}$, where $\nu$ is the number of communities contained in $\mathcal{C}$. If for any $i,j \in \{1, \cdots, \nu\}, i \ne j$, $C_i \cap C_j = \emptyset$, then $\mathcal{C}$ is a non-overlapping community division, otherwise it is a overlapping community division. This paper mainly studies the problem of non-overlapping community division.

We can also use a matrix to represent the community partitioning results. Let

\begin{equation}
    c_{ij}=\left\{\begin{array}{ll}
    1, & \text{ If } v_i \text{ belongs to } C_j  \\
      0, & \text{ otherwise}
\end{array}\right.
\end{equation}

\noindent then we obtain a matrix $(c_{ij})_{n \times \nu}$. It's not difficult to see that $\mathcal{C}$ and $(c_{ij})$ correspond one-to-one. So we also denote  $\mathcal{C}=(c_{ij})_{n \times \nu}$.

\subsection{Evaluation Indicator}

Below are three commonly used indicators for evaluating the effectiveness of community division.

{\bf Modularity}: Modularity is an indicator that measures the degree of closeness of connections within a community relative to those between communities. The calculation formula for modularity is as follows:

\begin{equation}
Q=\frac{1}{2m}\sum_{i,j}[a_{ij}-\frac{d(v_i)d(v_j)}{2m}]\delta(C_i, C_j)
\end{equation}

\noindent where $Q$ denotes the modularity, $m$ is the total number of edges in the network, $a_{ij}$ represents the connection strength between nodes $n_i$ and $n_j$, specifically referring to the element of the adjacency matrix $A$ in this paper. $d(v_i)$ and $d(v_j)$ respectively represent the degrees of nodes $v_i$ and $v_j$, $C_i$ and $C_j$ are the communities to which nodes $v_i$ and $v_j$ belong, and $\delta(C_i, C_j)$ equals to 1 when $C_i$ and $C_j$ are equal, otherwise equals to 0.

Modularity can be considered as the aggregate of the differences between the number of connected edges and the expectation value in each community. When the actual number of edges is higher than the expectation, the nodes in the graph are considered more concentrated.

Modularity does not require knowledge of the true community partitioning of the network. For Networks with ground-truth community structures, we can use Normalized Mutual Information ($N\!M\!I$) and $F1$-score to evaluate the quality of community partitioning.

{\bf $N\!M\!I$} \citep{Lancichinetti}: This indicator is used to measure the similarity between the detected community structures and the ground-truth community structures, whose definition is given as followings:

\begin{equation}
N\!M\!I(\mathcal{C}, \mathcal{C}') = 
\frac{-2 \sum_{C_i \in \mathcal{C}}\sum_{C_j \in \mathcal{C}'} \hat{C}_{ij}log(\frac{\hat{C}_{ij}n}{\hat{C}_{i\cdot}\hat{C}_{\cdot j}})}{\sum_{C_i \in \mathcal{C}}\hat{C}_{ij}log(\frac{\hat{C}_{i\cdot}}{n})+\sum_{C_j \in \mathcal{C}'}\hat{C}_{ij}log(\frac{\hat{C}_{\cdot j}}{n})}
\end{equation}

\noindent where $\mathcal{C}$ and $\mathcal{C}'$ are two matrices that represent the detected community structures and the ground-truth community structures, respectively. $\hat{C}$ is a confusion matrix, in which $\hat{C}_{ij}$ indicates the number of nodes in the network that belong to both community $C_i$ of $\mathcal{C}$ and community $C_j$ of $\mathcal{C}'$. $\hat{C}_{i\cdot}$ and $\hat{C}_{\cdot j}$ represent the sum of the $i$-th row and the $j$-th column of matrix $\hat{C}$, respectively.

The value of $N\!M\!I$ is between 0 and 1, and the higher the value, the better the performance of the algorithm.

{\bf $F1$-score} \citep{Yang}: This indicator also measures the closeness between the community structures discovered by the algorithm and the ground-truth community structures, which is given by the following formula:

\begin{equation}
  F1 = \frac{1}{2} \bigg(\frac{1}{|\mathcal{C}|}\sum_{\mathcal{C}_i\in \mathcal{C}}Fit(\mathcal{C}_i, \mathcal{C}'_{\rho(i)}) + \frac{1}{|\mathcal{C}'|}\sum_{\mathcal{C}'_j\in \mathcal{C}'}Fit(\mathcal{C}_{\rho'(j)}, \mathcal{C}'_j)\bigg)
\end{equation}

\noindent where $\mathcal{C}$ and $\mathcal{C}'$ are the detected and ground-truth communities, respectively.
$\rho(i)=argmax_j Fit(\mathcal{C}_i, \mathcal{C}'_j)$, $\rho'(j)=argmax_i Fit(\mathcal{C}_i, \mathcal{C}'_j)$, and $Fit(\mathcal{C}_i, \mathcal{C}'_j)$ is the harmonic mean of Precision and Recall, i.e.

\[F1(\mathcal{C}_i, \mathcal{C}'_j) = 2 \cdot \frac{{precision({\mathcal{C}_i}, {{\mathcal{C}'}_j}) \cdot recall({\mathcal{C}_i}, {{\mathcal{C}'}_j})}}{{precision({\mathcal{C}_i}, {{\mathcal{C}'}_j}) + recall({\mathcal{C}_i}, {{\mathcal{C}'}_j})}}\]

\noindent where

\[precision({\mathcal{C}_i}, {\mathcal{C}'_j}) = \frac{{\left| {{\mathcal{C}_i} \cap {{\mathcal{C}'}_j}} \right|}}{{\left| {{\mathcal{C}_i}} \right|}}\]

\noindent and

\[recall({\mathcal{C}_i}, {\mathcal{C}'_j}) = \frac{{\left| {{\mathcal{C}_i} \cap {{\mathcal{C}'}_j}} \right|}}{{\left| {{{\mathcal{C}'}_j}} \right|}}\]

The value range of $F1$-score is also between 0 and 1. The closer its value is to 1, the closer the result of community partitioning is to the ground-truth community structures. Conversely, the closer its value is to 0, the greater the difference between the result of community partitioning and the ground-truth community structures.

\subsection{Traditional Graph Convolutional Neural Network}

GCNNs learn the node attribute by combining convolution and aggregation operators with neighbors' information in the graph structure. The procedure of GCNN is shown in Figure \ref{GCNN}. The input layer consists of the adjacency matrix of the network and the attribute matrix of the nodes. The convolutional layer obtains new features of nodes by fusing the features of nodes and their neighbors. Through several layers of convolution operations, the attribute of node $n_i$ changes from $X_i$ to $Z_i$.

\begin{figurehere}
  \centering
  \includegraphics[height=4cm,width=8cm]{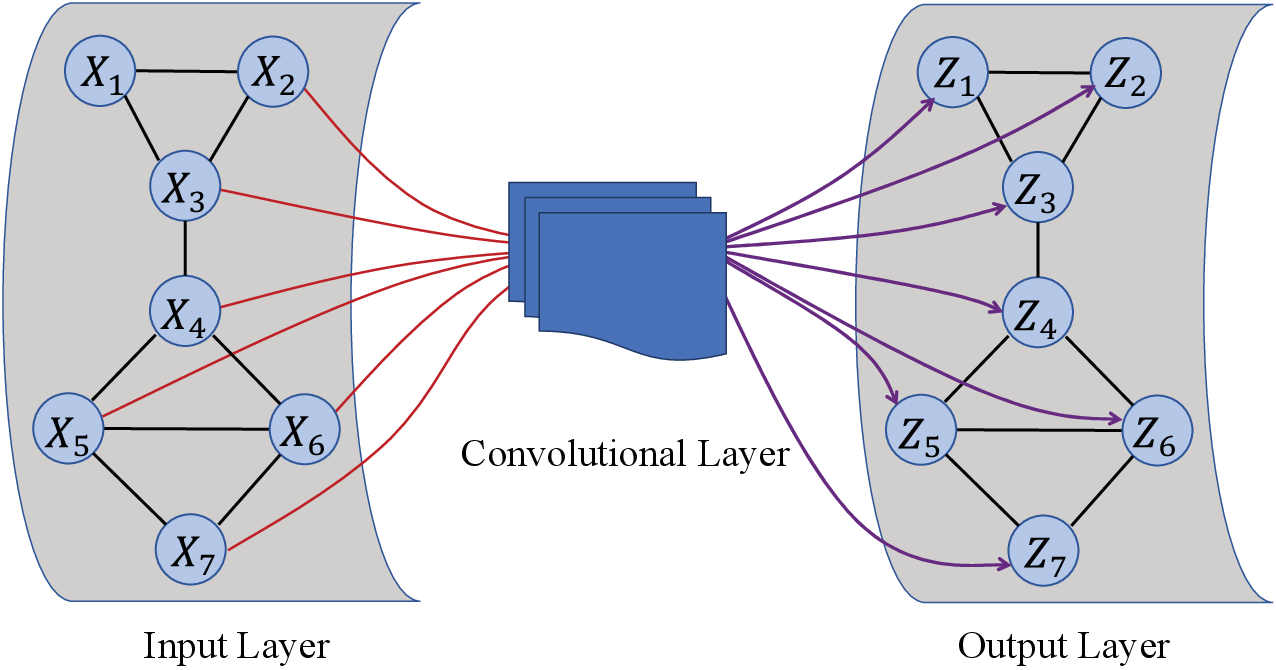}
 \caption{The structure of GCNN.}
 \label{GCNN}
\end{figurehere}

The specific form of the convolution operator is as follows:

\begin{equation}\label{Eq1}
H_{l+1}=\varphi(\tilde{D}^{-\frac{1}{2}}\tilde{A}\tilde{D}^{-\frac{1}{2}}H_l W_l)
\end{equation}

\noindent where $H_l$ is the input of layer $l+1$ and also the output of layer $l$. The initial input $H_0=X$. $\tilde{A}=A+I$, where $A$ is the adjacent matrix of the network and $I$ is the identify matrix with the same order with $A$. $\tilde{D}$ is the degree matrix of $\tilde{A}$, i.e. $\tilde{d}_{ii}=\sum_{j=1}^n \tilde{a}_{ij}$. $W_l$ is the weight matrix of level $l$, and $\varphi()$ is the activation function.

The output of GCNN is the non-negative community affiliation matrix $Z \in [0, 1]^{n \times \nu}$, where $Z_{ij}$ denotes the strength of node $v_i$
belonging community $C_j$. If we use a $L$-layer convolutional network, then

\begin{equation}\label{Eq2}
Z = H_{L} = \varphi(\tilde{D}^{-\frac{1}{2}}\tilde{A}\tilde{D}^{-\frac{1}{2}}H_{L-1} W_{L-1})\end{equation}

In formula (\ref{Eq1}), the convolution operation $\tilde{D}^{-\frac{1}{2}}\tilde{A}\tilde{D}^{-\frac{1}{2}}$ has weighted the attributes of nodes and their neighboring nodes based on their linking relationships. The function of a weight matrix $W_l$ is to add up the weighted attributes of all nodes and assign them to each node. 

To obtain ideal community partitioning results via GCNNs, it is necessary to train the weight matrix of each layer. Since the numbers of nodes of different networks varies, the orders of their weight matrices also varies. Additionally, even networks with the same number of nodes can have significant structural differences, resulting in different optimal weight matrices. As a result, each network needs tailored training. Consequently, applying conventional GCNNs to community partitioning problems involves considerable effort and presents substantial challenges.

To address the aforementioned issues, this paper proposes a community detection method based on graph convolution iterative algorithm (GCIS), which does not require weight training, but obtains the optimal community partition result by controlling the number of convolution operations through modularity indicator.

\section{Community Detection Method Based on Graph convolution Iteration Algorithm}

This section will elaboration the community detection method based on the graph convolution iteration algorithm.

\subsection{Overall Process}

The flow chart of the community detection algorithm proposed in this paper is shown in Figure \ref{Overall}.

\begin{figurehere}
  \centering
  \includegraphics[height=11cm,width=5cm]{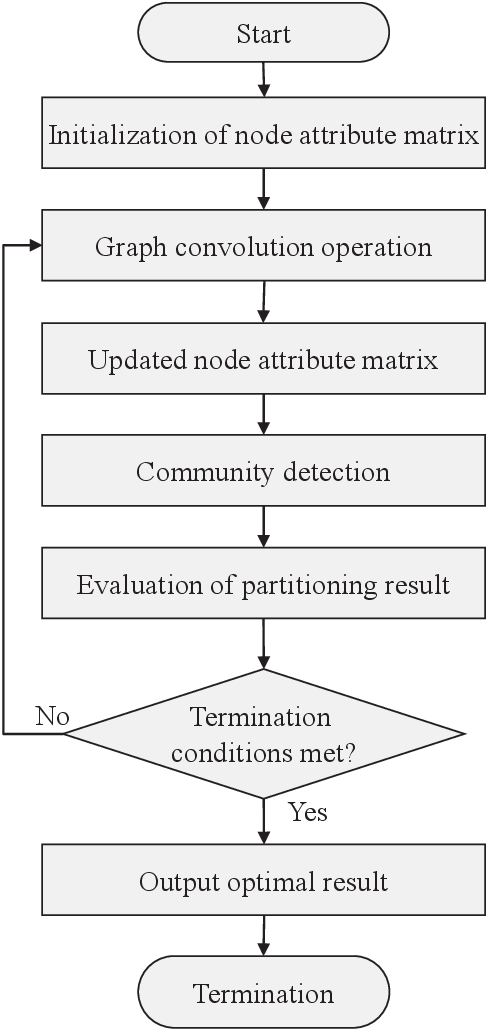}
 \caption{Flow chart of the algorithm.}
 \label{Overall}
\end{figurehere}

The main body of the algorithm consists of five parts: initialization of node attribute matrix, graph convolution operation, community partitioning, evaluation of partitioning result, and termination conditions.

\subsection{Initialization of Node Attribute Matrix}\label{Initialization}

Generally speaking, the closer a node is to a community center, the more likely it is to belong to the community. Therefore, we define its attributes based on the distance from the node to the community centers.

We obtain candidate community centers by randomly sampling from the vertex set. The sampling ratio can be determined based on the size of the network. Intuitively, this is a less reliable method due to the presence of significant randomness. But in reality, this strategy is beneficial for us to explore more possibilities.

Additionally, the candidate community centers may not necessarily be the final community centers. If all nodes have low proximity to a center, then this community will not contain any nodes, and thus this center will be automatically abandoned. So, the selection of community centers will not have a significant impact on the results of the algorithm. As long as the selected community center is an influential node in the community, other nodes in the community can generally be gathered together. But if all nodes are set as candidate community centers, the computational workload is too large. In addition, it is easy to cause inaccurate community partitioning results because of poor competition. For example, if two centers are very close, and their surrounding nodes are both very close to each other, the surrounding nodes may be divided up by them. Actually, these two communities should be merged. By random sampling, it is possible to effectively suppress unhealthy competition among nodes, which can actually benefit community division.

Suppose that $c_1, c_2, \cdots, c_{\alpha}$ are the candidate community centers. Generally speaking, $\alpha$ should be greater than $\nu$, the number of communities. However, we do not know the actual value of $\nu$, so $\alpha$ can be slightly larger to ensure that each community has some nodes selected. As long as a relatively influential node (not necessarily the optimal influential node) is selected in each community, it is possible to aggregate other nodes into one community. We also attempted to select community centers based on their influence, exploring various methods, but found that none were as effective as the random method.

Let $v_i$ be a node in the network and $dis(v_i, c_j)$ be the distance between $v_i$ and $c_j$, which is the length the shortest path between them in the network. Then, the proximity of $v_i$ and $c_j$ is defined as

\begin{equation}pro(v_i, c_j)=\frac{1}{dis(v_i, c_j)+1}\end{equation}

\noindent When $v_i=c_j$, then $dis(v_i, c_j)=0$, so their proximity is maximum, equal to 1. When there is no path connecting $v_i$ and $c_j$, then $dis(v, v_i)=\infty$, so $pro(v_i, c_j)=0$. The greater the proximity $pro(v_i, c_j)$, the greater the likelihood that node $v_i$ belongs to the community taking $c_j$ as center.

The proximity of $v_i$ to all community centers can be represented by a vector, called its attribute vector, denoted as $X_i$, i.e. $X_i=(x_{i1}, \cdots, x_{i\alpha})$. The attribute vectors of all vertices form a matrix, which is called the node's attribute matrix, denoted as $X$, i.e. $X=(x_{ij})_{n \times \alpha}$.

 Certainly, it is unreasonable to divide communities solely based on the distances between nodes and community centers. However, if a node's neighbors, or even the neighbors' neighbors, are very close to a community center, then the node is highly likely to belong to that community. And using graph convolution operations can precisely aggregate the attributes of nodes and their neighbors, thereby providing more accessible information for community partitioning.

\subsection{Graph Convolution Operation}\label{Operation}

As mentioned earlier, although the node attribute matrix provides a basis for community partitioning, directly relying on the attribute matrix for community partitioning would result in poor performance. So we use convolution operators to aggregate the attributes of neighboring nodes, in order to obtain better community partitioning results.

However, our graph convolution operation is significantly different from traditional GCNNs.

\begin{itemize}
  \item Firstly, the convolution matrix we use is the adjacency matrix $A$ of the network, not the transformed matrix $\tilde{A}=A+I$.
\end{itemize}

Traditional GCNNs assume that the diagonal elements of $A$ are all zero, meaning that only the attributes of neighboring nodes are aggregated when multiplying with the attribute matrix $X$, while the node's own attribute is overlooked. To address this issue, they incorporate the identity matrix $I$ into $A$ to create a new adjacency matrix.

Although the nodes' own attributes are not considered in the first convolution, they are included in the second convolution because they have been integrated into their neighbors' attributes during the first convolution. If we set all diagonal elements of the adjacency matrix to 1, it would amplify the node's own attributes, which could distort the proximity between nodes. Therefore, we will leave the adjacency matrix $A$ unchanged.

\begin{itemize}
  \item Secondly, our algorithm does not need to train the weight matrix.
\end{itemize}

In formula (\ref{Eq1}), the function of multiplying by the weight matrix $W_l$ is to weight and sum the attributes of all nodes by different coefficients, and then use the results as the new attributes of nodes. The convolution operation intends to aggregate the attributes of adjacent nodes, but $W_l$ aggregates all node regardless of whether they are adjacent or not, which is clearly problematic. In addition, even if the weight matrix can be trained to fuse valuable information, it will still repeat the previous convolution operation and greatly increase the workload. Therefore, we let $W_l = I$ in this paper.

\begin{itemize}
  \item Finally, we mainly control the number of convolution operators to obtain the optimal community partitioning result, rather than pre-set the number of convolutions.
\end{itemize}

The traditional method sets the number of convolution layers in advance and then adjusts the weights of different layers to achieve better partitioning results. Their iteration refers to the training process for weights. Nevertheless, each iteration of our algorithm refers to performing more convolution operations on node attributes. The best partition result will be recorded among all iterations. This process is similar to many iterative algorithms, such as genetic algorithms \cite{1989Genetic}.

In summary, our graph convolution operation is shown as follows: 

\begin{equation}\label{Eq3}
H_{l+1}=\varphi(D^{-\frac{1}{2}}AD^{-\frac{1}{2}}H_l )
\end{equation}

\noindent where $H_l$ is the input of layer $l+1$ and also the output of layer $l$. The initial input $H_0=X$. $A$ is the adjacent matrix of the network and $D$ is the degree matrix of $A$, i.e. $d_{ii}=\sum_{j=1}^n a_{ij}$. The activation function $\varphi()$ of hidden layers is the function ReLU, and that of the output layer is the function Softmax.

Because $A$ is an unormalized matrix, and multiplying it with the attribute matrix $H_l$ will change the original distribution of the attributes. So we will standardize $A$ to obtain a symmetric and normalized matrix $\hat{A} = D^{-\frac{1}{2}}AD^{-\frac{1}{2}}$. Because $D$ can be obtained by $A$, $A$ is one of the inputs of the GCNN. The other input is the node attribute $X$.

The output of GCNN is the non-negative community affiliation matrix $Z \in [0, 1]^{n \times \alpha}$, where $Z_{ij}$ denotes the strength of node $v_i$ belonging community $C_j$.

\subsection{Community Partitioning}\label{Detection}

We can obtain the final community partitioning result based on the output of the algorithm. Suppose that the output corresponding node $n_i$ is $Z_i = \{z_{i1}, z_{i2}, \cdots, z_{i\alpha}\}$, where $z_{ij}$ is the probability that $n_i$ belonging to community $C_j$. Let $k=arg max \{z_{i1}, z_{i2}, \cdots, z_{i\alpha}\}$, then $n_i$ will be divided into community $C_k$. 

It should be noted that we divide communities based on maximum probability rule instead of traditional probability based random method. This is because each value in $Z_i$ is very small for large networkds and the differences between different elements are more small, partitioning based on probability will bring a lot of uncertainty.

In addition, during the convolution process, some communities may be annexed by others. That is to say, the probability of a community center belonging to its own community may be lower than the probability of belonging to another community. For example, when two community centers are relatively close and their links are very intense, their respective communities are likely to merge into a larger community. Suppose that $\nu$ nonempty communities are obtained in the end. 

Since we do not train the weight matrix, we will continuously increase the number of convolutions in order to obtain the optimal partitioning result. Therefore, our algorithm can be seen as an optimization process, with the output being a community partitioning result.

\subsection{Evaluation of Partitioning Result}\label{Evaluation}

In general, we cannot guarantee that the network has ground-truth community structures. Therefore, we use modularity to evaluate the results of community partitioning. The purpose of the algorithm is to achieve maximum modularity through convolution operations.

\subsection{Termination Conditions}\label{Termination}

Like most iterative algorithms, we can set different termination conditions for the algorithm proposed in this paper. For example, we can give a threshold for the total number of iterations, or the algorithm can be terminated when the evaluation metric, i.e. modularity, does not increase after several iterations.

\subsection{Supplementary Explanation}

Although the random method is used to determine community centers, which has a certain degree of uncertainty, the effectiveness of community partitioning will not be greatly affected. After convolution, the proximity between nodes within the community will be higher than that between nodes of different communities. 

In addition, we can also eliminate the impact of random factors by sampling multiple times, and the increased computational complexity will not be significant. According to formula (\ref{Eq4}), the previous matrix operations are all the same, and different sampling results will only affect the node attribute matrix $X$. 

We can first use all nodes as candidate community centers to obtain the node attribute matrix $X$. Suppose that the candidate community centers by random sampling is $X'$. Let

\begin{equation}
 \gamma_i=\left\{
 \begin{array}{ll}
  1, & \text{if node } n_i \text{ is selected} \\
  0, & \text{otherwise}
\end{array}
\right.
\href{}{}\end{equation}

\noindent then we can obtain a vector $\Gamma=(\gamma_1, \cdots, \gamma_n)$. Let $S=diag(\Gamma)$ is the matrix with elements in $\Gamma$ as the diagonal. Then

\[X' = XS\]

According to formula (\ref{Eq6}), we have
\[Z'=\hat{A}_lX'=\hat{A}_lXS=ZS\]

From the above process, it can be seen that if we sample partial nodes as community centers, we only need to sample the corresponding columns of the final output matrix. Therefore, we can perform multiple samples without increasing a significant amount of computational complexity, as the previous convolution process is completely identical.

\section{Case Study}

We chose a small network to analyze the implementation process of the algorithm proposed in this paper.

\begin{figurehere}
  \centering
\includegraphics[height=5cm,width=6cm]{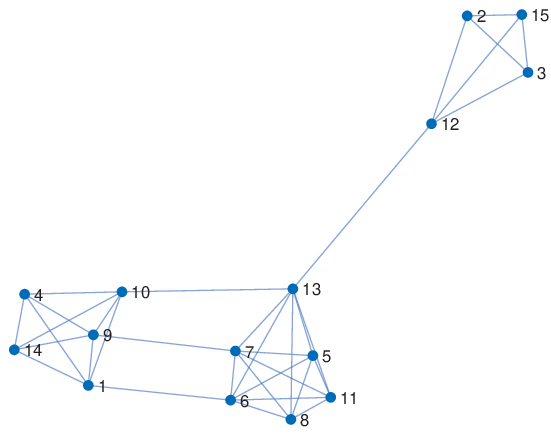}
 \caption{Example network.}
 \label{network}
\end{figurehere}

The structure of the example network is shown in Figure \ref{network}. This network contains 15 nodes and 33 edges. We can see that it has 3 very clear community structures, namely $\{2, 3, 12, 15\}$, $\{1, 4, 9, 10, 14\}$ and $\{5, 6, 7, 8, 11, 13\}$. Next, we will illustrate how the method proposed in this paper detect communities by graph convolution operations.

Firstly, the node attribute matrix will be constructed using the method provided in Section \ref{Initialization}. We will use two different methods to obtain the candidate community centers. The first one does not sample but treat all nodes as candidate community centers. While the second method will use the method of random sampling to select one-third of the nodes as candidate community centers.

{\bf Case 1: Treat all nodes as candidate community centers}

We first do not sample but treat all nodes as candidate community centers to obtain the node attribute matrix $X$ as formula (\ref{Eq4}).

\begin{figure*}[t]\tiny
\begin{equation}\label{Eq4} 
X=\left(
  \begin{array}{ccccccccccccccc}
{\bf 1.0000} & 0.2000 & 0.2000 & 0.5000 & 0.3333 & 0.5000 & 0.3333 & 0.3333 & 0.5000 & 0.5000 & 0.3333 & 0.2500 & 0.3333 & 0.5000 & 0.2000 \\
0.2000 & {\bf 1.0000} & 0.5000 & 0.2000 & 0.2500 & 0.2500 & 0.2500 & 0.2500 & 0.2000 & 0.2500 & 0.2500 & 0.5000 & 0.3333 & 0.2000 & 0.5000 \\
0.2000 & 0.5000 & {\bf 1.0000} & 0.2000 & 0.2500 & 0.2500 & 0.2500 & 0.2500 & 0.2000 & 0.2500 & 0.2500 & 0.5000 & 0.3333 & 0.2000 & 0.5000 \\
0.5000 & 0.2000 & 0.2000 & {\bf 1.0000} & 0.2500 & 0.3333 & 0.3333 & 0.2500 & 0.5000 & 0.5000 & 0.2500 & 0.2500 & 0.3333 & 0.5000 & 0.2000 \\
0.3333 & 0.2500 & 0.2500 & 0.2500 & {\bf 1.0000} & 0.5000 & 0.5000 & 0.5000 & 0.3333 & 0.3333 & 0.5000 & 0.3333 & 0.5000 & 0.2500 & 0.2500 \\
0.5000 & 0.2500 & 0.2500 & 0.3333 & 0.5000 & {\bf 1.0000} & 0.5000 & 0.5000 & 0.3333 & 0.3333 & 0.5000 & 0.3333 & 0.5000 & 0.3333 & 0.2500 \\
0.3333 & 0.2500 & 0.2500 & 0.3333 & 0.5000 & 0.5000 & {\bf 1.0000} & 0.5000 & 0.5000 & 0.3333 & 0.5000 & 0.3333 & 0.5000 & 0.3333 & 0.2500 \\
0.3333 & 0.2500 & 0.2500 & 0.2500 & 0.5000 & 0.5000 & 0.5000 & {\bf 1.0000} & 0.3333 & 0.3333 & 0.5000 & 0.3333 & 0.5000 & 0.2500 & 0.2500 \\
0.5000 & 0.2000 & 0.2000 & 0.5000 & 0.3333 & 0.3333 & 0.5000 & 0.3333 & {\bf 1.0000} & 0.5000 & 0.3333 & 0.2500 & 0.3333 & 0.5000 & 0.2000 \\
0.5000 & 0.2500 & 0.2500 & 0.5000 & 0.3333 & 0.3333 & 0.3333 & 0.3333 & 0.5000 & {\bf 1.0000} & 0.3333 & 0.3333 & 0.5000 & 0.5000 & 0.2500 \\
0.3333 & 0.2500 & 0.2500 & 0.2500 & 0.5000 & 0.5000 & 0.5000 & 0.5000 & 0.3333 & 0.3333 & {\bf 1.0000} & 0.3333 & 0.5000 & 0.2500 & 0.2500 \\
0.2500 & 0.5000 & 0.5000 & 0.2500 & 0.3333 & 0.3333 & 0.3333 & 0.3333 & 0.2500 & 0.3333 & 0.3333 & {\bf 1.0000} & 0.5000 & 0.2500 & 0.5000 \\
0.3333 & 0.3333 & 0.3333 & 0.3333 & 0.5000 & 0.5000 & 0.5000 & 0.5000 & 0.3333 & 0.5000 & 0.5000 & 0.5000 & {\bf 1.0000} & 0.3333 & 0.3333 \\
0.5000 & 0.2000 & 0.2000 & 0.5000 & 0.2500 & 0.3333 & 0.3333 & 0.2500 & 0.5000 & 0.5000 & 0.2500 & 0.2500 & 0.3333 & {\bf 1.0000} & 0.2000 \\
0.2000 & 0.5000 & 0.5000 & 0.2000 & 0.2500 & 0.2500 & 0.2500 & 0.2500 & 0.2000 & 0.2500 & 0.2500 & 0.5000 & 0.3333 & 0.2000 & {\bf 1.0000} 
  \end{array}
\right)\end{equation}\end{figure*}

Next, we will use the method provided in Section \ref{Operation} to perform convolution operations on the attribute matrix $X$. After each convolution, we can obtain the corresponding community partitioning results and calculate the modularity using the methods provided in Sections \ref{Detection} and \ref{Evaluation}, respectively. The termination condition of the algorithm is that the modularity has not increased for two consecutive iterations. Finally, the algorithm ends in the 7th generation, and the optimal partitioning result is obtained in the 5th generation.

The output of the fifth convolution operation is shown in formula (\ref{Eq5}). We can see that after 5 convolution operations, the node attribute matrix has undergone significant changes. This is because, through convolution operations, the attributes of nodes not only include distance information from themselves to the candidate community centers, but also integrate attributes of their neighbors.

\begin{figure*}[t]\tiny
\begin{equation}\label{Eq5}
H_5=\left(
  \begin{array}{ccccccccccccccc}
0.5049 & 0.2426 & 0.2426 & 0.4800 & 0.4011 & 0.4510 & 0.4500 & 0.4011 & 0.5062 & {\bf 0.5073} & 0.4011 & 0.3183 & 0.4524 & 0.4800 & 0.2426 \\
0.2227 & 0.5104 & 0.5124 & 0.2146 & 0.2871 & 0.2882 & 0.2882 & 0.2871 & 0.2227 & 0.2739 & 0.2871 & {\bf 0.5260} & 0.3816 & 0.2146 & 0.5124 \\
0.2227 & 0.5124 & 0.5104 & 0.2146 & 0.2871 & 0.2882 & 0.2882 & 0.2871 & 0.2227 & 0.2739 & 0.2871 & {\bf 0.5260} & 0.3816 & 0.2146 & 0.5124 \\
0.4686 & 0.2131 & 0.2131 & 0.4492 & 0.3459 & 0.3947 & 0.3947 & 0.3459 & 0.4686 & {\bf 0.4693} & 0.3459 & 0.2788 & 0.3962 & 0.4497 & 0.2131 \\
0.4015 & 0.2761 & 0.2761 & 0.3482 & 0.4818 & 0.4977 & 0.4977 & 0.4819 & 0.4015 & 0.4059 & 0.4819 & 0.3613 & {\bf 0.5069} & 0.3482 & 0.2761 \\
0.4586 & 0.2963 & 0.2963 & 0.4036 & 0.5144 & 0.5364 & 0.5374 & 0.5144 & 0.4574 & 0.4616 & 0.5144 & 0.3879 & {\bf 0.5462} & 0.4036 & 0.2963 \\
0.4574 & 0.2963 & 0.2963 & 0.4036 & 0.5144 & 0.5374 & 0.5364 & 0.5144 & 0.4586 & 0.4616 & 0.5144 & 0.3879 & {\bf 0.5462} & 0.4036 & 0.2963 \\
0.4015 & 0.2761 & 0.2761 & 0.3482 & 0.4819 & 0.4977 & 0.4977 & 0.4818 & 0.4015 & 0.4059 & 0.4819 & 0.3613 & {\bf 0.5069} & 0.3482 & 0.2761 \\
0.5062 & 0.2426 & 0.2426 & 0.4800 & 0.4011 & 0.4500 & 0.4510 & 0.4011 & 0.5049 & {\bf 0.5073} & 0.4011 & 0.3183 & 0.4524 & 0.4800 & 0.2426 \\
0.5035 & 0.2507 & 0.2507 & 0.4782 & 0.3989 & 0.4474 & 0.4474 & 0.3989 & 0.5035 & {\bf 0.5046} & 0.3989 & 0.3230 & 0.4533 & 0.4782 & 0.2507 \\
0.4015 & 0.2761 & 0.2761 & 0.3482 & 0.4819 & 0.4977 & 0.4977 & 0.4819 & 0.4015 & 0.4059 & 0.4818 & 0.3613 & {\bf 0.5069} & 0.3482 & 0.2761 \\
0.2769 & 0.5301 & 0.5301 & 0.2611 & 0.3473 & 0.3514 & 0.3514 & 0.3473 & 0.2769 & 0.3246 & 0.3473 & {\bf 0.5537} & 0.4435 & 0.2611 & 0.5301 \\
0.4769 & 0.3706 & 0.3706 & 0.4236 & 0.5435 & 0.5655 & 0.5655 & 0.5435 & 0.4769 & 0.4910 & 0.5435 & 0.4639 & {\bf 0.5880} & 0.4236 & 0.3706 \\
0.4686 & 0.2131 & 0.2131 & 0.4497 & 0.3459 & 0.3947 & 0.3947 & 0.3459 & 0.4686 & {\bf 0.4693} & 0.3459 & 0.2788 & 0.3962 & 0.4492 & 0.2131 \\
0.2227 & 0.5124 & 0.5124 & 0.2146 & 0.2871 & 0.2882 & 0.2882 & 0.2871 & 0.2227 & 0.2739 & 0.2871 & {\bf 0.5260} & 0.3816 & 0.2146 & 0.5104 
 \end{array}
\right)\end{equation}\end{figure*}

The bold mathematics in formula (\ref{Eq5}) represents the maximum element in each row. We can see that all the maximum elements are concentrated in columns 10, 12, and 13. According to the community partitioning rules given in Section \ref{Detection}, all nodes will be divided into three communities, i.e. $\{2, 3, 12, 15\}$, $\{1, 4, 9, 10, 14\}$ and $\{5, 6, 7, 8, 11, 13\}$. The community partitioning result is completely consistent with the expectation. Although we consider all nodes as candidate community centers, there are only three real community centers in the end, i.e. nodes 10, 12, and 13. From Figure \ref{network}, it can be seen that these three nodes are also very appropriate as community centers. Firstly, nodes 12 and 13 both have the largest degrees in their respective communities. Although node 10 has the same degree as node 9 and node 1, it is connected to another community center: node 13. Therefore, its influence is the greatest in its community. 

{\bf Case 2: Sample one third of nodes as candidate community centers}

Next, we will sample one third of nodes as candidate community centers. Through random sampling, the nodes we selected are 3, 5, 7, 9 and 14. The node attribute matrix $X'$ is obtained according Section \ref{Initialization}, shown as formula (\ref{Eq6}). We can see that, $X'$ in formulas (\ref{Eq6}) is composed of the 3th, 5th, 7th, 9th and 14th columns of the matrix $X$ in formula (\ref{Eq4}). Let 

\[S=diag(0, 0, 1, 0, 1, 0, 1, 0, 1, 0, 0, 0, 1, 0)\]

\noindent then, $X'=XS$.

\begin{equation}\label{Eq6}\tiny
X' =\left(
\begin{array}{ccccc}
0.2000 & 0.3333 & 0.3333 & 0.5000 & 0.5000 \\
0.5000 & 0.2500 & 0.2500 & 0.2000 & 0.2000 \\
{\bf 1.0000} & 0.2500 & 0.2500 & 0.2000 & 0.2000 \\
0.2000 & 0.2500 & 0.3333 & 0.5000 & 0.5000 \\
0.2500 & {\bf 1.0000} & 0.5000 & 0.3333 & 0.2500 \\
0.2500 & 0.5000 & 0.5000 & 0.3333 & 0.3333 \\
0.2500 & 0.5000 & {\bf 1.0000} & 0.5000 & 0.3333 \\
0.2500 & 0.5000 & 0.5000 & 0.3333 & 0.2500 \\
0.2000 & 0.3333 & 0.5000 & {\bf 1.0000} & 0.5000 \\
0.2500 & 0.3333 & 0.3333 & 0.5000 & 0.5000 \\
0.2500 & 0.5000 & 0.5000 & 0.3333 & 0.2500 \\
0.5000 & 0.3333 & 0.3333 & 0.2500 & 0.2500 \\
0.3333 & 0.5000 & 0.5000 & 0.3333 & 0.3333 \\
0.2000 & 0.2500 & 0.3333 & 0.5000 & {\bf 1.0000} \\
0.5000 & 0.2500 & 0.2500 & 0.2000 & 0.2000 
 \end{array}
\right)\end{equation}

Similarly, using the method presented in this paper, the optimal community partitioning results are obtained in the 3rd iteration. The convolution result of the 3rd iteration is shown in formula (\ref{Eq7}).

\begin{equation}\label{Eq7}\tiny
H'_3=\left(
  \begin{array}{ccccc}
0.2316 & 0.3787 & 0.4365 & {\bf 0.5367} & 0.5182 \\
{\bf 0.5569} & 0.2758 & 0.2754 & 0.2116 & 0.2079 \\
{\bf 0.5384} & 0.2758 & 0.2754 & 0.2116 & 0.2079 \\
0.2037 & 0.3204 & 0.3776 & {\bf 0.5037} & 0.4938 \\
0.2636 & 0.5048 & {\bf 0.5173} & 0.3804 & 0.3198 \\
0.2827 & 0.5364 & {\bf 0.5539} & 0.4420 & 0.3808 \\
0.2827 & 0.5364 & {\bf 0.5468} & 0.4506 & 0.3808 \\
0.2636 & 0.5088 & {\bf 0.5173} & 0.3804 & 0.3198 \\
0.2316 & 0.3787 & 0.4443 & {\bf 0.5273} & 0.5182 \\
0.2420 & 0.3752 & 0.4325 & {\bf 0.5332} & 0.5157 \\
0.2636 & 0.5088 & {\bf 0.5173} & 0.3804 & 0.3198 \\
{\bf 0.5694} & 0.3365 & 0.3391 & 0.2631 & 0.2512 \\
0.3615 & 0.5645 & {\bf 0.5808} & 0.4612 & 0.4012 \\
0.2037 & 0.3204 & {\bf 0.3776} & 0.5037 & 0.4860 \\
{\bf 0.5569} & 0.2758 & 0.2754 & 0.2116 & 0.2079 
 \end{array}
\right)\end{equation}

From formula (\ref{Eq7}), it can be seen that the optimal values for each row are also concentrated in three columns, based on which the entire network can be divided into three communities, i.e. $\{2, 3, 12, 15\}$, $\{1, 4, 9, 10, 14\}$ and $\{5, 6, 7, 8, 11, 13\}$, which is the same with the result of the first method.

In fact, if the result of the third convolution in Case 1 is $H_3$, then $H_3'=H_3S$.

Through this example, it can be seen that although the candidate community centers we selected are random, we can still obtain the true community partitioning results. This is because the link strength between nodes in the same community is significantly higher than that between nodes in different communities. As long as we choose a node in the community as the center, we can aggregate the nodes in that community into that community. 

To distinguish between these two different strategies, we label the method that does not perform random sampling as GCI, while the method that performs sampling as GCIS. We plot the modularity, $N\!M\!I$, and $F1$-score of the community partitioning results obtained in each iteration for these two methods, as shown in Figure \ref{example_curve}.

\begin{figurehere}
   \centering
   \subfigure[GCI]
   {
       \includegraphics[width=0.22\textwidth]{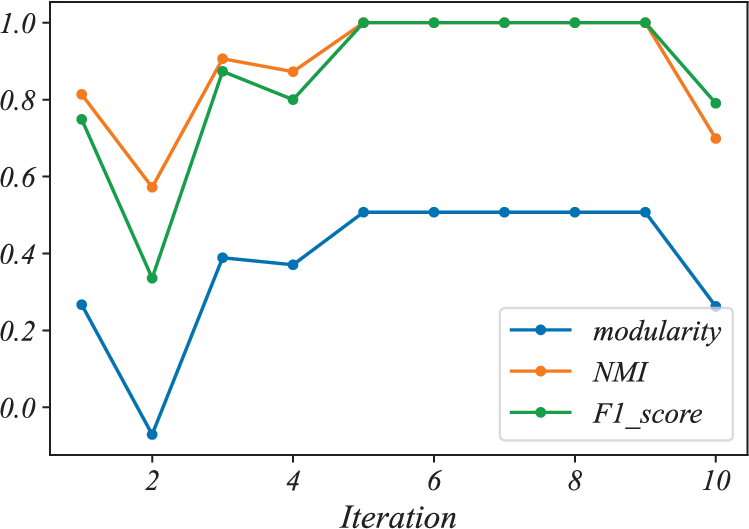}
    }
   \subfigure[GCIS]
   {
       \includegraphics[width=0.22\textwidth]{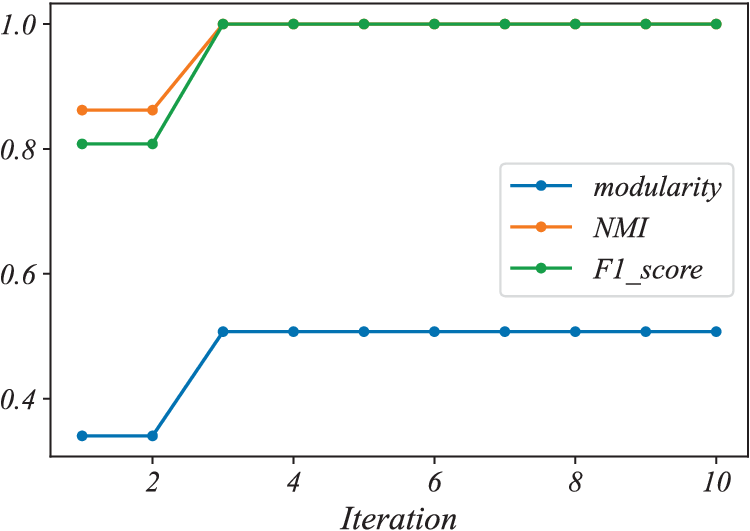}
    }
   \caption{Changing curves of different indicators for the example network.}
   \label{example_curve}
\end{figurehere}

From Figure \ref{example_curve}, we can see that (1) As the number of iterations (or the number of convolutional layers) increases, all indicators first continuously increase, reach the highest value, then tend to stabilize, and finally decrease at some point; (2) The trends in modularity, NMI, and F1-score are completely consistent. So using modularity as an optimization indicator can find the ground-truth community structure; (3) Both methods obtained real community partitioning results, but random sampling method makes the algorithm converge to the optimal value faster. This is because, by sampling, it is possible to eliminate negative competition among nodes with similar community aggregation power, thereby obtaining better partitioning results. In addition, sampling can reduce the size of the node attribute matrix, thereby reducing the computational complexity required for each convolution.

\section{Experiment}

This section validates the effectiveness of the proposed
algorithm through a series of experiments. First, the experimental subjects are listed. Then, the baseline methods are introduced. Finally, the experimental results are provided and analyzed.

\subsection{Experimental Subjects}

To verify the effectiveness of our method, we selected two kinds of networks for experiments. 

The first class of networks are random networks (RNs). Firstly, we give the total number of nodes in the network, and then set the number of communities in advance. When generating edges, nodes within the same community are connected with a high probability, while nodes between different communities are connected with a low probability.

The number of nodes in the first set of random networks is set to 100, and the number of communities is set to 3. The number of nodes in each community varies according to certain rules. In addition, in order to compare the impact of different connection probabilities on algorithm performance, the connection probabilities are taken as different values. The information of this group of networks is listed in Table \ref{tab1}, in which $P_1$ is the connecting probability of nodes in the same communities and $P_2$ is the connecting probability of nodes in different communities.

\begin{tablehere}
\footnotesize\centering
\caption{Information of First Group of Random Networks.}
\label{tab1}       
\renewcommand\arraystretch{1.0}
\begin{tabular}{llll}
\toprule
Network & Vertices of communities & $P_1$ & $P_2$\\\hline
RN$_1$ & 10/10/80 & 0.9/0.8/0.7/0.6 &0.05\\
RN$_2$ & 10/20/70 & 0.9/0.8/0.7/0.6 &0.05\\
RN$_3$ & 10/30/60 & 0.9/0.8/0.7/0.6 &0.05\\
RN$_4$ & 10/40/50 & 0.9/0.8/0.7/0.6 &0.05\\
RN$_5$ & 20/20/60 & 0.9/0.8/0.7/0.6 &0.05\\
RN$_6$ & 20/30/50 & 0.9/0.8/0.7/0.6 &0.05\\
RN$_7$ & 20/40/40 & 0.9/0.8/0.7/0.6 &0.05\\
RN$_8$ & 30/30/40 & 0.9/0.8/0.7/0.6 &0.05
\\\bottomrule
 \end{tabular}
\end{tablehere}

The numbers of nodes in the second set of random networks increase from 200 to 5000, with a connection probability of 0.8 for nodes in the same community and 0.05 for nodes from different communities. The number of nodes included in each community is determined using the random method. The specific information is listed in Table \ref{tab2}.

\begin{tablehere}
\footnotesize\centering
\caption{Information of Second Group of Random Networks.}
\label{tab2}      
\renewcommand\arraystretch{1.0}
\begin{tabular}{lll}
\toprule
Network & Vertices of communities & Communities \\\hline
RN$_9$ & 200 & 8 \\
RN$_{10}$ & 300 & 8 \\
RN$_{11}$ & 400 & 8 \\
RN$_{12}$ & 500 & 8 \\
RN$_{13}$ & 1000 & 10 \\
RN$_{14}$ & 2000 & 10 \\
RN$_{15}$ & 3000 & 10 \\
RN$_{16}$ & 4000 & 10 \\
RN$_{17}$ & 5000 & 10
\\\bottomrule
 \end{tabular}
\end{tablehere}

We also chose four real-world networks for experiments, whose detailed information are listed in Table \ref{tab3}. All of these networks have ground-truth community structures.

\begin{table*}
\footnotesize\centering
\caption{Information of real-world networks.}
\label{tab3}       
\renewcommand\arraystretch{1.0}
\begin{tabular}{lllllll}
\toprule
Network & \#Nodes & \#Edges & \#Communities & Average node degree & $CC$ & Reference\\\hline
Karate & 34  &78 & 2 & 4.5882 & 0.5706 & \citep{2002Community}\\
Dolphin & 62 & 159 & 2 & 5.1290 & 0.2559 & \citep{2003The}\\
Polbooks & 105 & 441 & 3 & 8.4000 & 0.4875 & \citep{2002Community}  \\ 
Football & 115 & 613 & 12 & 10.6609 & 0.4032 & \citep{Informationzhang}\\
DBLP & 317080 & 1049866 & 13477 & 6.6222 & 0.6324 & \citep{snapnets} \\
Amazon & 334863 & 925872 & 75149 & 5.5299 & 0.3967 & \citep{snapnets} 
\\\bottomrule
\end{tabular}
\end{table*}

\subsection{Baseline Methods}

To verify the performance of the proposed method, five classic community detection algorithms are selected for comparison: KL algorithm \citep{1970An}, GN algorithm \citep{2004Finding}, Walktrap algorithm \citep{2005Computing}, Louvain algorithm \citep{2008Fast} and COPRA algorithm \citep{2009Finding}. In addition, we also chose two deep learning based methods for comparison, namely Wu's method \citep{WuStudy} and Al-Andoli's method \citep{NASSERALANDOLI202294}.

KL algorithm is a binary method that divides a known network into two communities of known size, and it is a greedy algorithm. Its main idea is to define a function gain $Q$ for network partitioning. 

GN algorithm divides network nodes naturally into densely connected subgroups. The algorithm has two definitive features: firstly, it continuously removes edges from the network, dividing them into multiple communities and identifying the removed edges using some measures. Secondly, it is crucial to recalculate these measures after each removal.

Walktrap algorithm calculates similarity between vertices (and communities) based on random walks. Compared with previous methods, this distance measurement method has higher computational efficiency and can effectively capture information about community structure.

Louvain algorithm is a modularity based graph algorithm model. Unlike ordinary modularity based and modularity based gains, this algorithm is fast and performs well in clustering graphs with fewer edges and edges.

COPRA algorithm is a community detection algorithm based on GCNNs. This algorithm and the algorithm in this paper are both based on graph convolution operation, but the specific implementation method is completely different.

Wu's method provides an unsupervised community detection algorithm based on graph convolutional network model for application.

Al-Andoli's method is a parallel deep learning-based community detection method in large complex networks with a hybrid backpropagation-particle swarm optimization. 

In addition, to further validate the effectiveness of random sampling, we still use two strategies to select the initial community centers, corresponding to method GCI and GCIS, respectively.

\subsection{Experimental Results and Analysis}

\subsubsection{Impact of Connecting Probability on Algorithm Performance}

To verify the practicality of GCIS in networks with different connection strengths, we applied it to the first set of random networks, where all other settings were the same, except that the connection probabilities between nodes in the same community are 0.9, 0.8, 0.7, and 0.6, respectively. 

Finnaly, GCIS obtains the maximum $N\!M\!I$ and $F1$-score value 1 for all networks, which indicates that our method can accurately detect all communities. The maximum modularity of each network is different, which is determined by the structure of the network itself. In addition, as the connecting probability changes, the modularity of the network does not change very significantly. Nevertheless, as the number of nodes included in each community changes, the modularity undergoes significant changes. When there is a significant difference in the number of nodes included in each community, the modularity of the network is relatively low. While When the number of nodes in each community does not differ significantly, the modularity of the network is relatively high.

\subsubsection{Impact of Convolution Times on Algorithm Performance}

\begin{figure*}
   \centering
   \subfigure[RN$_1$]
   {
       \includegraphics[width=0.17\textwidth]{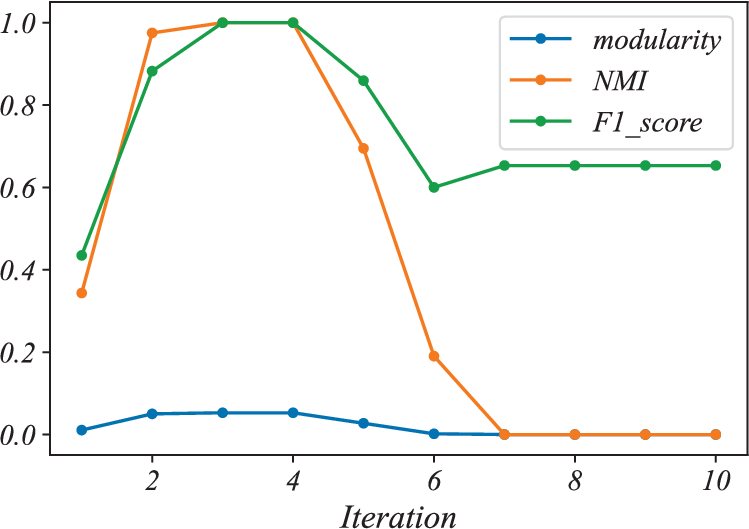}
    }
   \subfigure[RN$_2$]
   {
       \includegraphics[width=0.17\textwidth]{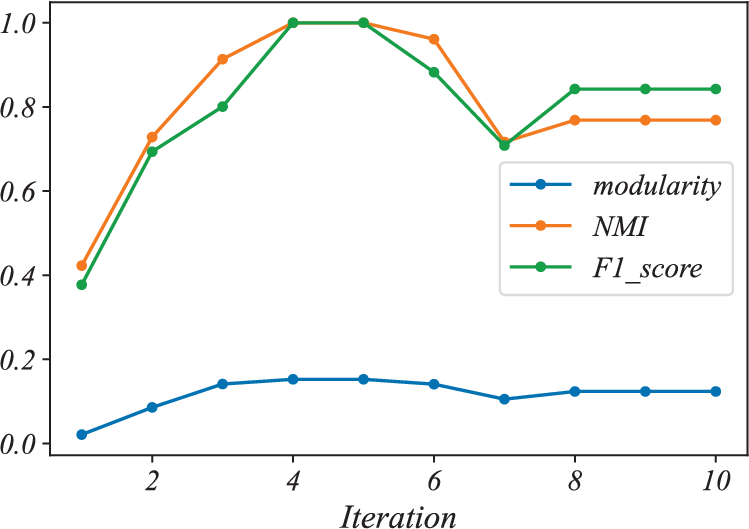}
    }
   \subfigure[RN$_3$]
   {
       \includegraphics[width=0.17\textwidth]{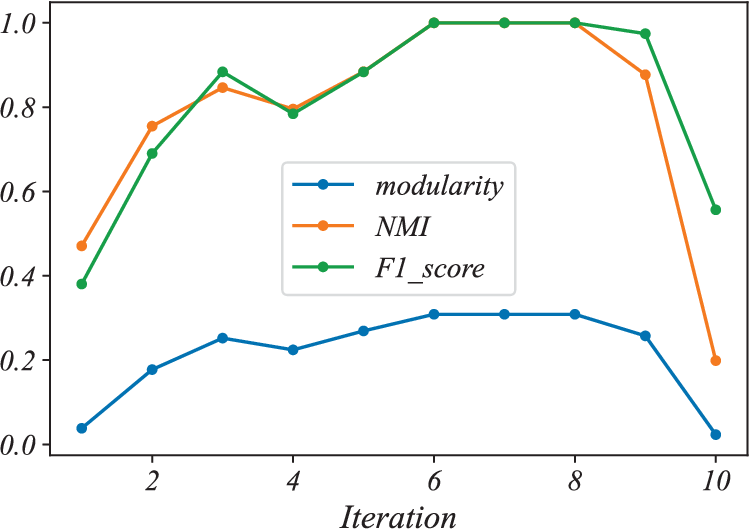}
    }
   \subfigure[RN$_4$]
   {
       \includegraphics[width=0.17\textwidth]{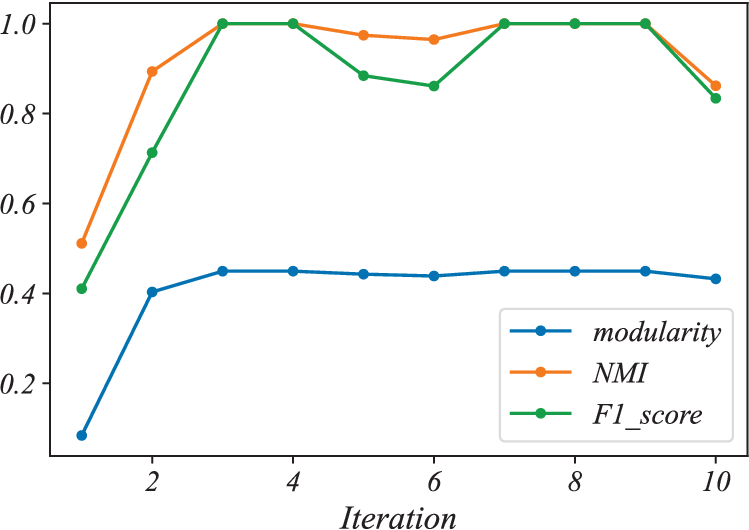}
    }
   \subfigure[RN$_5$]
   {
       \includegraphics[width=0.17\textwidth]{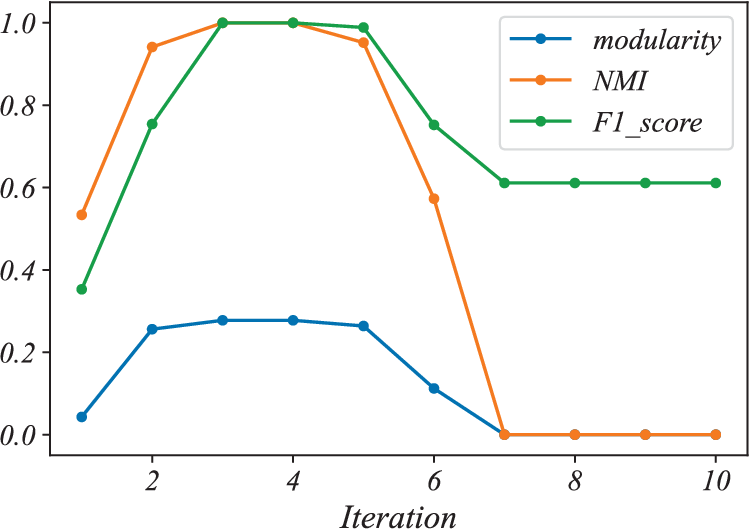}
    }
   \subfigure[RN$_6$]
   {
       \includegraphics[width=0.17\textwidth]{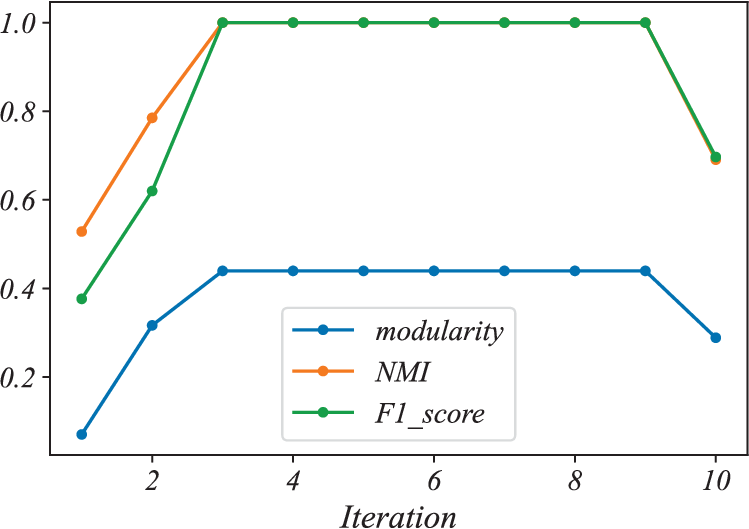}
    }
   \subfigure[RN$_7$]
   {
       \includegraphics[width=0.17\textwidth]{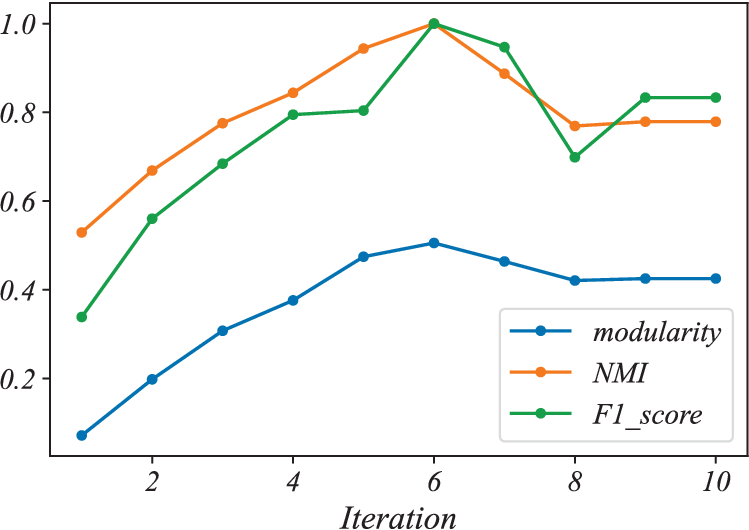}
    }
   \subfigure[RN$_8$]
   {
       \includegraphics[width=0.17\textwidth]{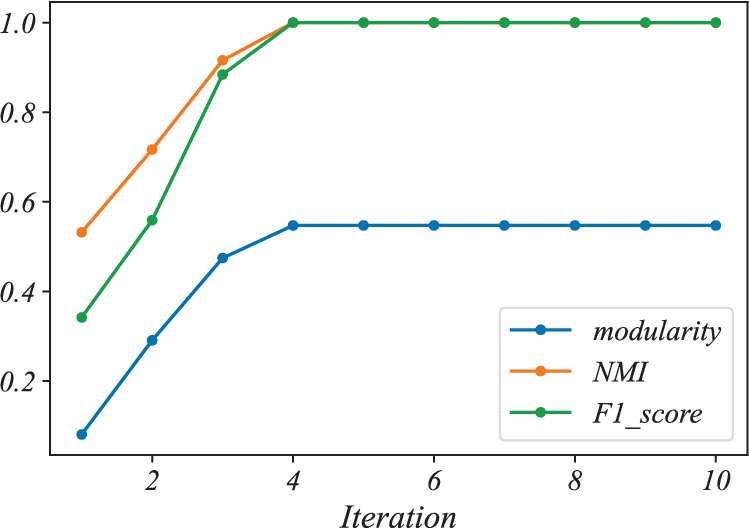}
    }
   \subfigure[RN$_9$]
   {
       \includegraphics[width=0.17\textwidth]{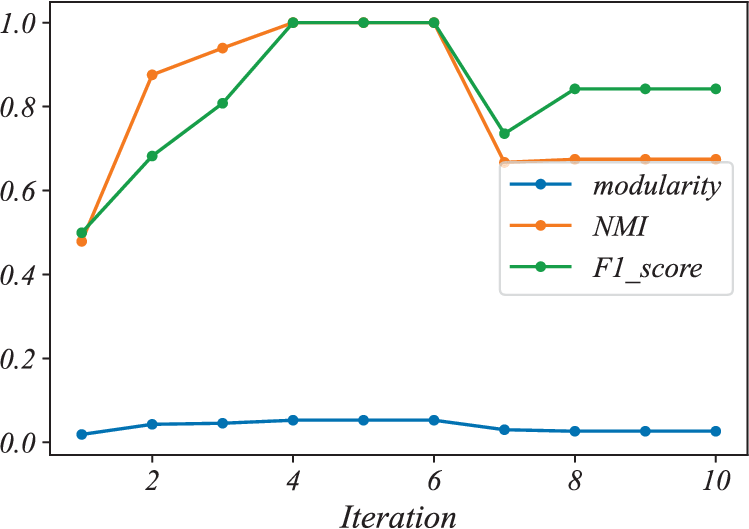}
    }
   \subfigure[RN$_{10}$]
   {
       \includegraphics[width=0.17\textwidth]{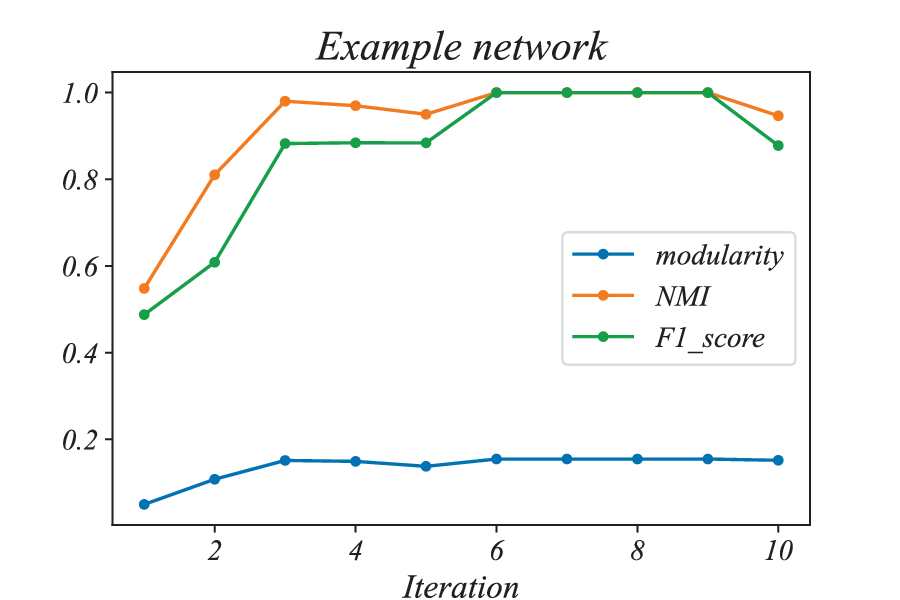}
    }
   \caption{Changing curves of different indicators for the example network.}
   \label{Evaluation indicator}
\end{figure*}

To observe the relationship between the number of convolutions and community partitioning results, we varied the number of convolutions from 1 to 10, and then calculated the modularity, $N\!M\!I$, and $F1$-score of the obtained partitioning results. Because different connection probabilities have little impact on the method proposed in this paper, we set the connection probability of nodes in the same community to 0.8. The final result is shown in Figure \ref{Evaluation indicator}.

From Figures \ref{Evaluation indicator}, it could be seen that, (1) As the number of convolutions increases, the modularity of the community partitioning results obtained by each network will rapidly increase at the beginning, and then the climbing speed will decrease, and even repeat. After reaching a maximum value, it will decrease. (2) Almost all networks can reach the highest modularity level in the first five generations. (3) As the number of convolutions increases, the trend of the three indicators is completely consistent, either increasing or decreasing at the same time, which indicates that for this highly structured random network, using these three indicators to evaluate the effectiveness of network partitioning is equivalent.

The above results indicate that the efficiency of the algorithm proposed in this paper is high, and it does not require too much convolution operation to achieve very ideal community partitioning results. In addition, it is reasonable to use modularity to evaluate the results of community partitioning, as the changing curve of the three indicators are basically consistent.

\subsubsection{Comparison with Benchmark Methods}

In order to fully verify the performance of the algorithm proposed in this paper, we conducted three sets of comparative experiments.

\begin{itemize}
  \item The first group of comparative experiments:
\end{itemize}

\begin{figure*}
  \centering
  \includegraphics[height=7cm,width=15cm]{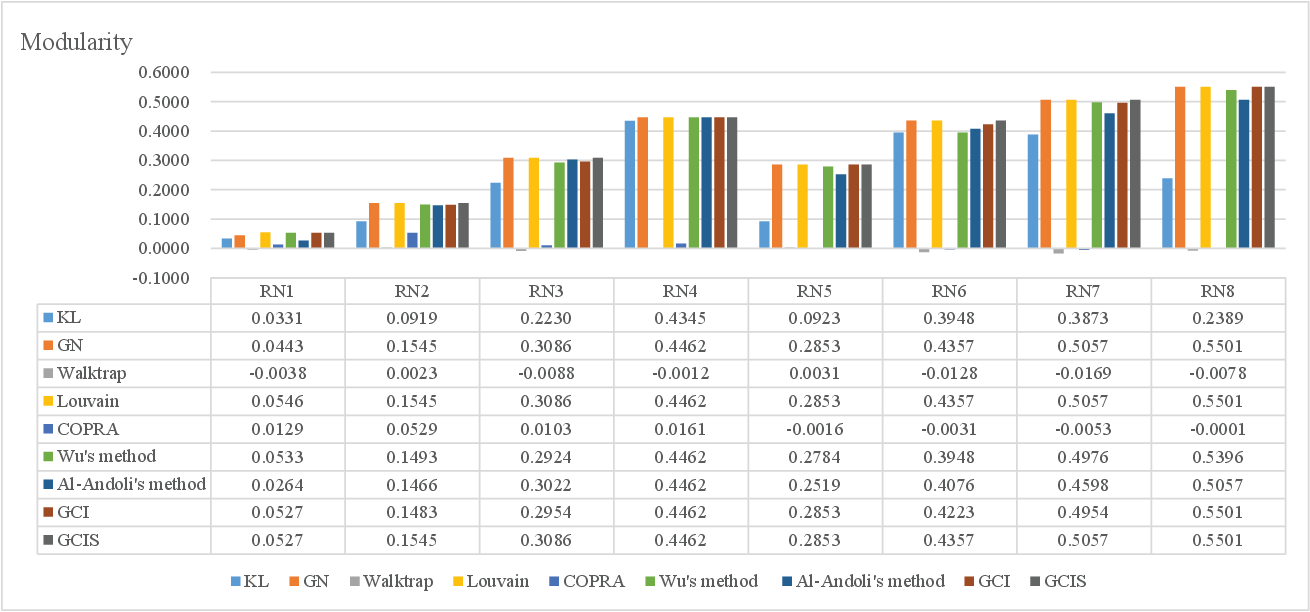}
 \caption{Modularities obtained by different methods in the first set of random networks.}
 \label{Modilarity-rand1}
\end{figure*}

\begin{figure*}
  \centering
  \includegraphics[height=7cm,width=15cm]{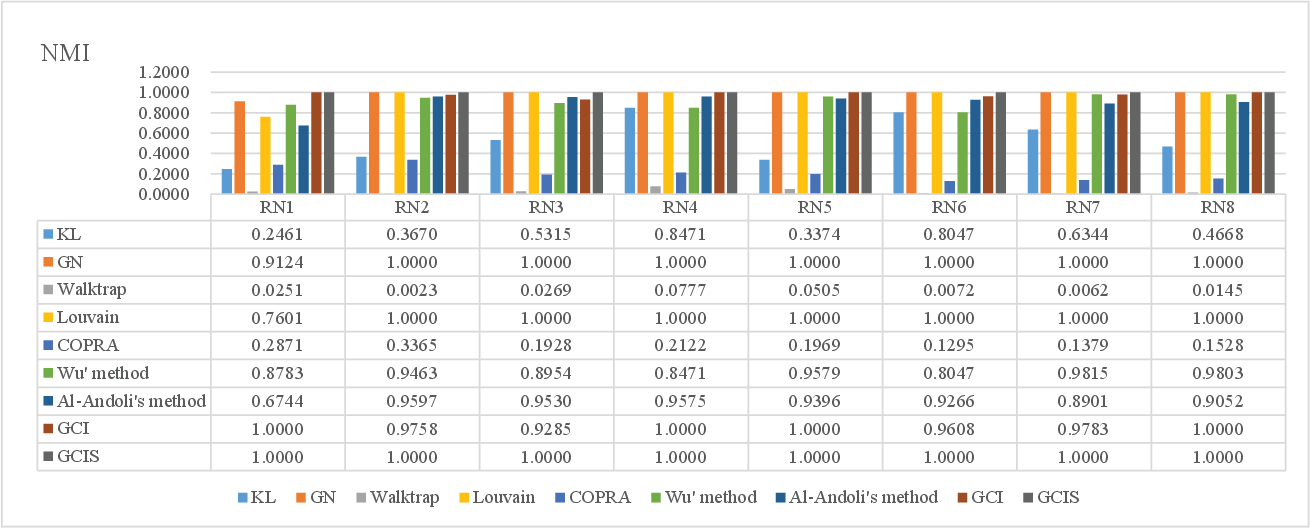}
 \caption{$N\!M\!I$ obtained by different methods in the first set of random networks.}
 \label{NMI-rand1}
\end{figure*}

\begin{figure*}
  \centering
  \includegraphics[height=7cm,width=15cm]{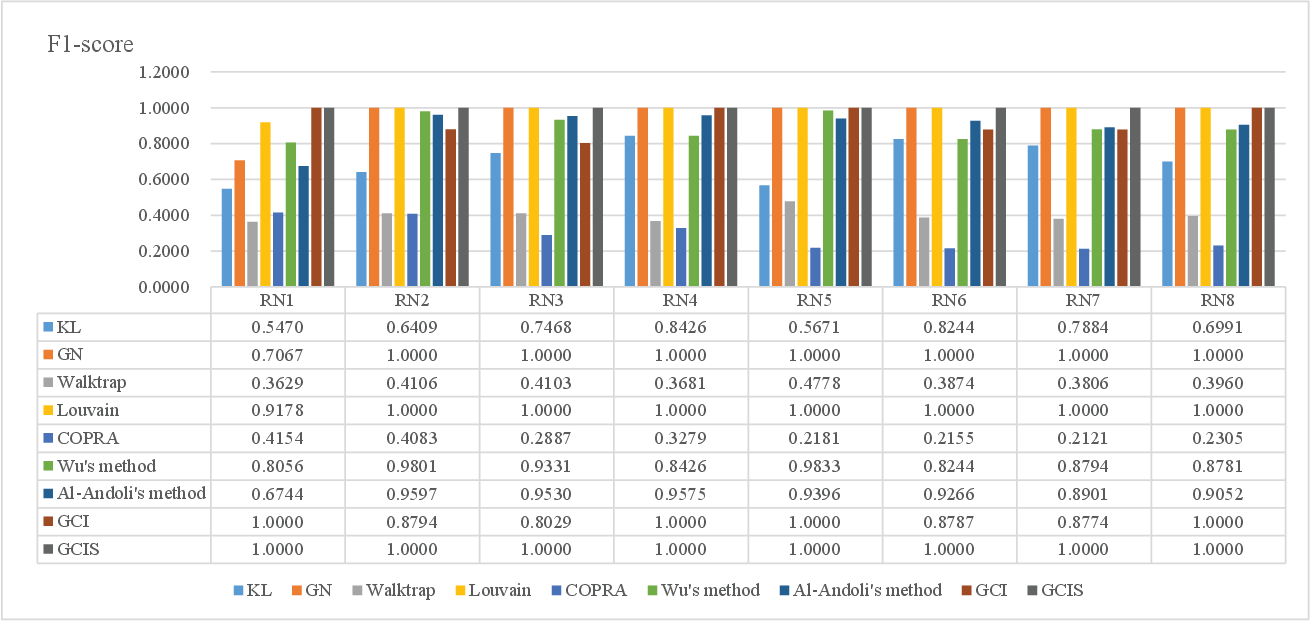}
 \caption{$F1$-score obtained by different methods in the first set of random networks.}
 \label{F1-rand1}
\end{figure*}

This group of experiments takes the random networks in Table \ref{tab1} as test objects. The probability of node connections within the same community is 0.8, and the probability of node connections between communities is 0.05. We used different methods to detect communities for these networks, and calculate the values of various evaluation indicators. The final comparison results are shown in Figures \ref{Modilarity-rand1}, \ref{NMI-rand1}, \ref{F1-rand1} and Table \ref{Time}.

\begin{table*}
\footnotesize\centering
\caption{Runtime of different methods in the first set of random networks(Unit: $s$).}
\label{Time}       
\renewcommand\arraystretch{1.0}
\begin{tabular}{llllllllll}
\toprule
Network & KL & GN & Walktrap & Louvain & COPRA & Wu's method & Al-Andoli's method  & GIC & GICS\\\hline
RN$_1$ & 0.00951  & 63.87396  & 0.16524  & 0.02394  & 0.42082  & 0.17342  & 0.21552  & 0.02888  & 0.01492 \\
RN$_2$ & 0.01118  & 41.20214  & 0.14582  & 0.01157  & 0.31400  & 0.18235  & 0.25244  & 0.03488  & 0.01193 \\
RN$_3$ & 0.00716  & 29.84501  & 0.13154  & 0.01056  & 0.31729  & 0.15363  & 0.22533  & 0.02194  & 0.00794 \\
RN$_4$ & 0.00461  & 42.66162  & 0.13839  & 0.01033  & 0.24192  & 0.15736  & 0.21543  & 0.01791  & 0.01094 \\
RN$_5$ & 0.00507  & 25.64171  & 0.23670  & 0.01118  & 0.25126  & 0.20426  & 0.26348  & 0.05884  & 0.01193 \\
RN$_6$ & 0.00500  & 20.10053  & 0.14078  & 0.00800  & 0.21270  & 0.14737  & 0.22735  & 0.01991  & 0.00795 \\
RN$_7$ & 0.00458  & 19.13733  & 0.15591  & 0.00750  & 0.19829  & 0.16724  & 0.25353  & 0.01891  & 0.00894 \\
RN$_8$ & 0.00399  & 15.89032  & 0.13127  & 0.01152  & 0.22368  & 0.16433  & 0.23743  & 0.02290  & 0.01991 
\\\bottomrule
 \end{tabular}
\end{table*}

From Figures \ref{Modilarity-rand1}, \ref{NMI-rand1}, \ref{F1-rand1} and Table \ref{Time}, it can be seen that, (1) GCIS achieves the best $N\!M\!I$ and $F1$-score in all networks, i.e. 1, that means the detected communities are completely consistent with the ground-truth community structures. The obtained modularity is also the highest for each network. (2) GCI only obtains the ground-truth community structures in four networks, which indicates that the strategy of sampling nodes is effective. (3) Among other comparison methods, the GN algorithm and Louvain algorithm have the best partitioning performance. However, neither of these two algorithms doesn't obtain the ground-truth community partition for Network 1. The number of nodes in the three communities of RN$_1$ is 10, 10, and 80, respectively. The result indicates that these two algorithms have poor performance in dealing with imbalanced networks. (4) Although GCI is not as effective as GCIS, it is still significantly better than KL, Walktrap, and COPRA. (5) Although the GN algorithm has achieved similar partitioning results as the Louvain algorithm, its runtime is much higher than other methods. The time required for Walktrap and COPRA is significantly reduced compared to GN, it is slightly higher than other three methods. The KL algorithm has the highest efficiency, but it divides all networks into two communities, so the effect is also relatively poor. The time spent on our algorithm and Louvain algorithm is comparable. (6) Both deep learning based methods have achieved good partitioning results, but there is still a certain gap compared to our method.

The above comparison results indicate that our algorithm not only has good partitioning performance, but also high efficiency.

\begin{itemize}
  \item The second group of comparative experiments:
\end{itemize}

To further validate the performance of the proposed method, we applied it to a larger scale random network as shown in Table \ref{tab2}. Based on the results of the first group of experiments, we only selected the Louvain algorithm,  for comparison, because its performance is similar to our algorithm. Additionally, we still use both GCI and GCIS strategies separately. The final comparison results are listed in Table \ref{Compare2}.

\begin{table*}\small
\footnotesize\centering
\caption{ comparison results of different methods for second comparative experiment.}
\label{Compare2}       
\renewcommand\arraystretch{1.0}
\begin{tabular}{ccccccccccccc}
\toprule
Method & Indicator &  RN$_9$  &  RN$_{10}$  &RN$_{11}$  &  RN$_{12}$  &  RN$_{13}$  &  RN$_{14}$  &  RN$_{15}$  &  RN$_{16}$  &  RN$_{17}$  \\\hline
\multirow[]{4}{*}{Louvain} & Modularity & 0.5237  & 0.6977  & 0.6898  & 0.6545  & 0.6880  & 0.7203  & 0.6840  & 0.7192  & 0.7056 \\
& $N\!M\!I$ & 0.95315 & 0.9540  & 0.9818  & 0.9831  & 0.9784  & 1.0000  & 0.9584  & 0.9936  & 0.9752 \\
& $F1$-score & 0.8665  & 0.9406  & 0.9411  & 0.9399  & 0.9021  & 1.0000  & 0.8015  & 0.9524  & 0.9078 \\
& Time($s$) & 0.0329  & 0.0379  & 0.0798  & 0.1147  & 0.4189  & 1.3963  & 4.5728  & 7.3991  & 10.1319 \\\hline
\multirow[]{4}{*}{Wu's method} & Modularity & 0.5129  & 0.6948  & 0.6771  & 0.6522  & 0.6627  & 0.7203  & 0.6828  & 0.7179  & 0.7064 \\
& $N\!M\!I$ & 0.9827  & 0.9797  & 0.9443  & 0.9913  & 0.9615  & 1.0000  & 0.9743  & 0.9950  & 0.9917 \\
& $F1$-score & 0.9390  & 0.9070  & 0.9225  & 0.9921  & 0.8550  & 1.0000  & 0.7613  & 0.9159  & 0.9195 \\
& Time($s$) & 0.0696  & 0.1456  & 0.2941  & 0.4119  & 1.7853  & 6.3583  & 15.4523  & 28.4724  & 44.7641 \\\hline
\multirow[]{4}{*}{Al-Andoli's method} & Modularity & 0.4553  & 0.7198  & 0.6536  & 0.6535  & 0.6807  & 0.7203  & 0.6823  & 0.7149  & 0.7066 \\
& $N\!M\!I$ & 0.7406  & 0.9940  & 0.9135  & 0.9980  & 0.9533  & 1.0000  & 0.9470  & 0.9683  & 0.9970 \\
& $F1$-score & 0.6803  & 0.9472  & 0.8792  & 0.9468  & 0.8155  & 1.0000  & 0.8042  & 0.9306  & 0.9495 \\
& Time($s$) & 0.0793  & 0.1389  & 0.2216  & 0.4535  & 1.7875  & 6.1570  & 14.3356  & 27.8731  & 48.0365 \\\hline
\multirow[]{4}{*}{GCI} & Modularity & 0.3391  & 0.6831  & 0.6794  & 0.6276  & 0.6609  & 0.7168  & 0.6805  & 0.7191  & 0.7068 \\
& $N\!M\!I$ & 0.8162  & 0.8914  & 0.9681  & 0.9727  & 0.8906  & 0.9974  & 0.9665  & 1.0000  & 0.9991 \\
& $F1$-score & 0.6841  & 0.8588  & 0.8656  & 0.8382  & 0.8029  & 0.9172  & 0.5889  & 1.0000  & 0.9589 \\
& Time($s$) & 0.1566  & 0.3441  & 0.3231  & 0.4792  & 2.1362  & 5.6520  & 10.1897  & 14.9041  & 24.3304 \\\hline
\multirow[]{4}{*}{GCIS} & Modularity & 0.5228  & 0.7278  & 0.6898  & 0.6544  & 0.6858  & 0.7203  & 0.6839  & 0.7191  & 0.7069 \\
& $N\!M\!I$ & 1.0000  & 1.0000  & 1.0000  & 1.0000  & 0.9918  & 1.0000  & 0.9566  & 1.0000  & 0.9951 \\
& $F1$-score & 1.0000  & 1.0000  & 1.0000  & 1.0000  & 0.8777  & 1.0000  & 0.8013  & 1.0000  & 0.9550 \\
& Time($s$) & 0.0478  & 0.0748  & 0.2333  & 0.1735  & 0.7849  & 1.7588  & 9.0719  & 33.5064  & 20.6158 
\\\bottomrule
 \end{tabular}
\end{table*}

From Table \ref{Compare2}, we can see that, (1) as the network size increases, GCIS still exhibits very stable performance. Out of 9 networks, 6 have obtained the true network structure. Among the three networks that do not obtain the true structure, except for network 15, the $N\!M\!I$ and $F1$-score values of the other two networks are both better than those of the Louvain algorithm. The result of Network 15 is only slightly inferior to the Louvain algorithm. (2) The Louvain algorithm and GCI both obtain the true partition of only one network.(3) For the runtime, GCIS is slightly higher than Louvain algorithm, but the difference is not very significant.

\begin{itemize}
  \item The third group of comparative experiments:
\end{itemize}

The third group of experiments uses real-world networks from Table \ref{tab3} as test subjects. The modularity, NMI, $F1$-score, number of communities, and running time of the networks obtained by different methods are compared.

\begin{figure*}
   \centering
   \subfigure[Karate]
   {
       \includegraphics[width=0.35\textwidth]{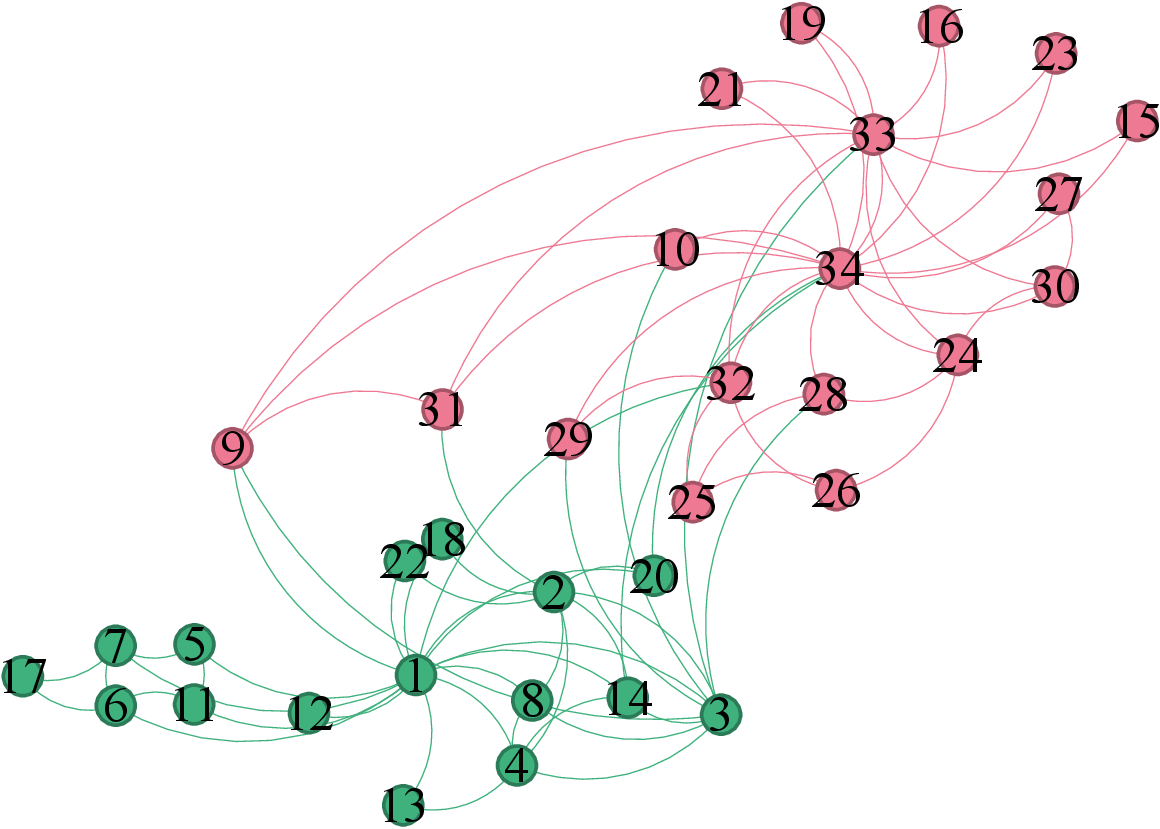}
    }
   \subfigure[Dolphin]
   {
       \includegraphics[width=0.35\textwidth]{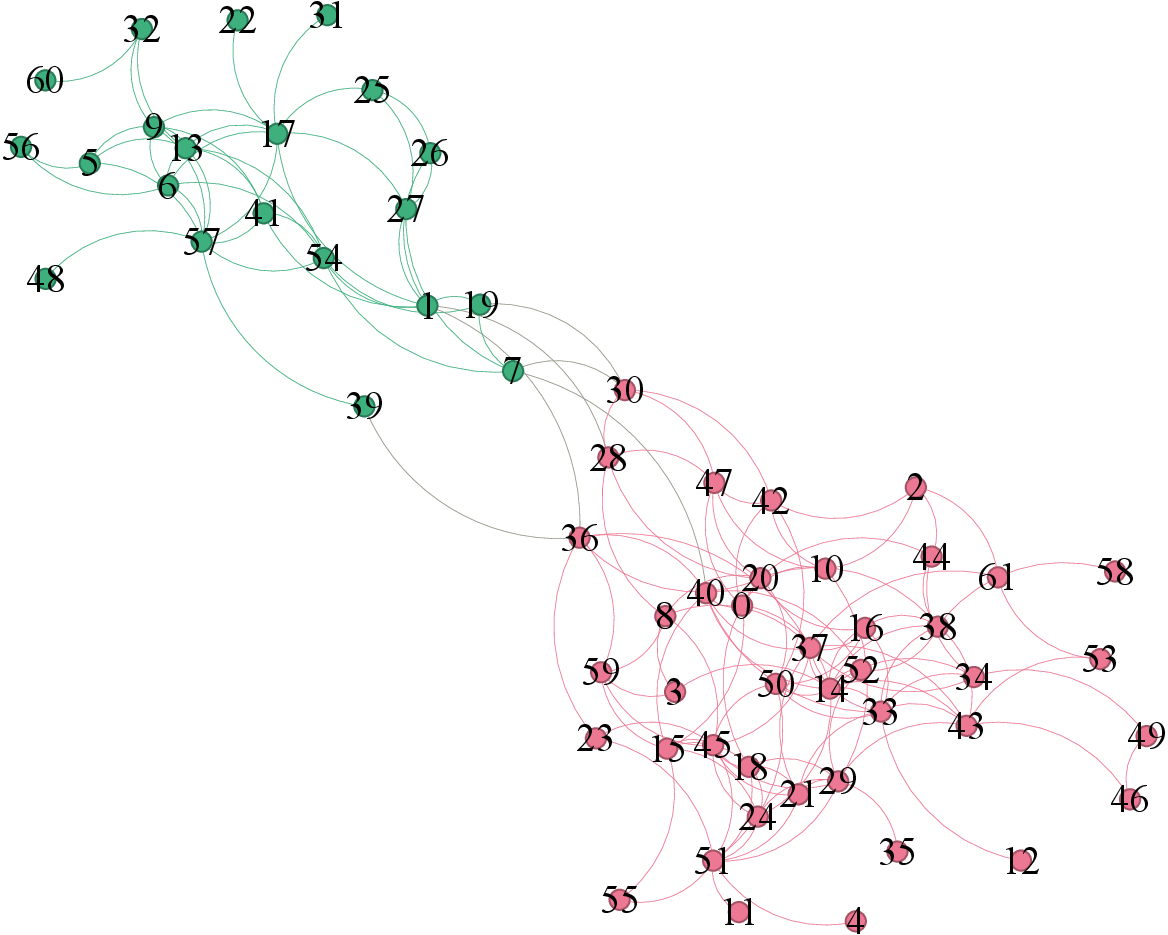}
    }
   \subfigure[Polbooks]{%
       \includegraphics[width=0.35\textwidth]{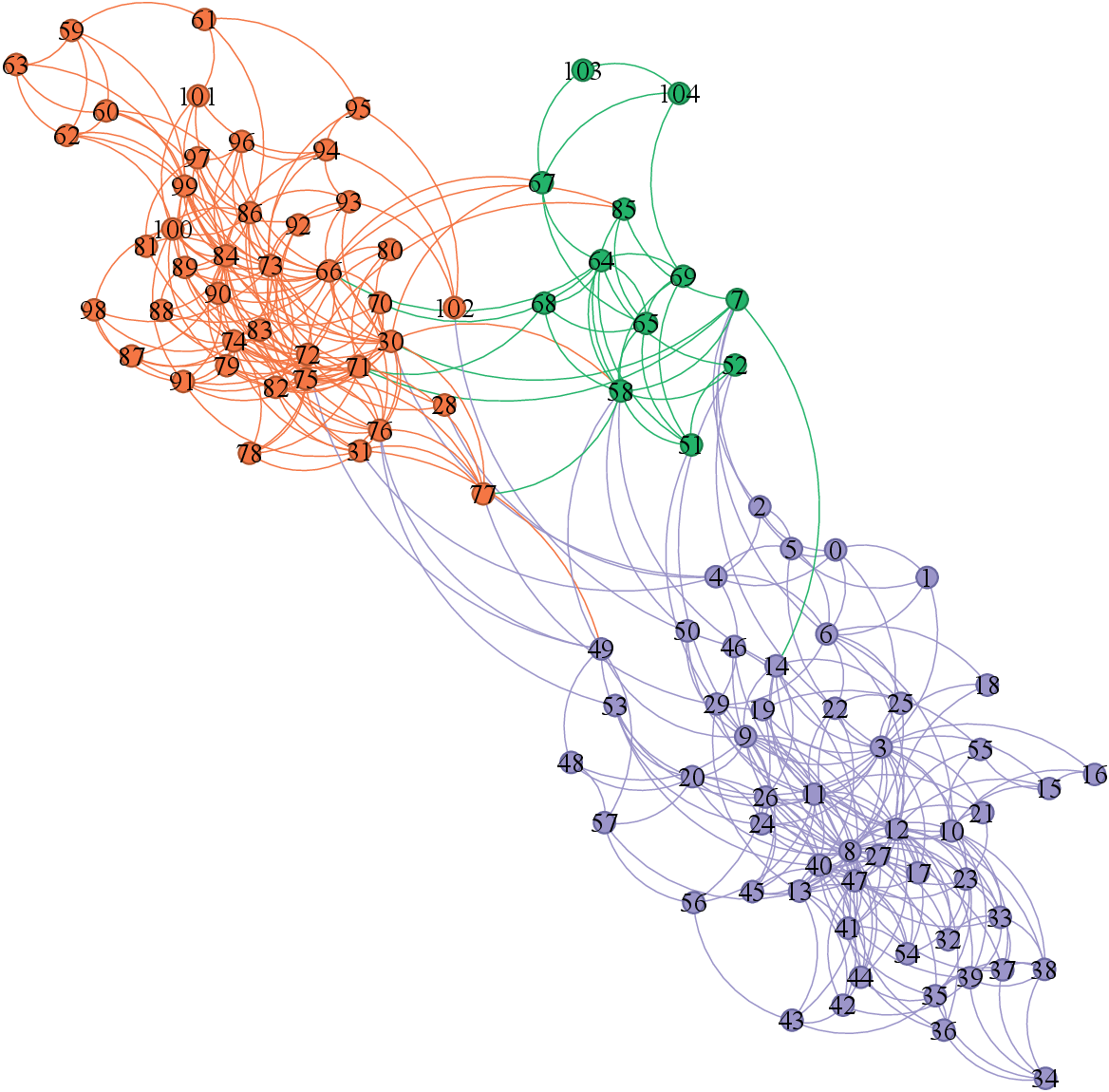}
   }
   \subfigure[Football]{%
       \includegraphics[width=0.35\textwidth]{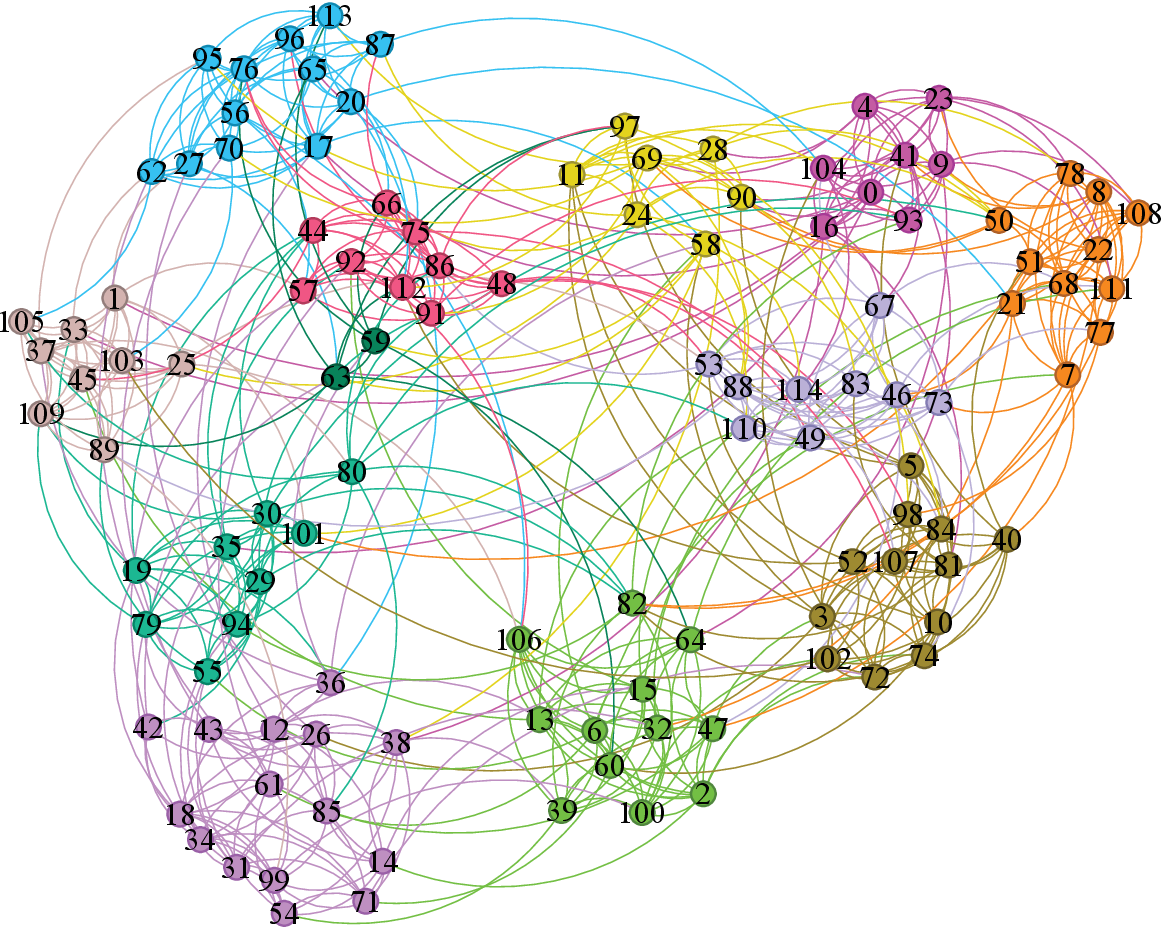}
   }
   \caption{The community detection results of the real-world networks using our method.}
   \label{Communities_real}
\end{figure*}

First, we use the method in this paper to divide the selected networks into communities, and visualized partitioning results for these four small-scale networks are shown in Figure \ref{Communities_real}, in which different colors represent different communities. 

From Figure \ref{Communities_real}, it can be seen that our method can obtain clear community structures for all networks.

\begin{figure*}
  \centering
  \includegraphics[height=7cm,width=15cm]{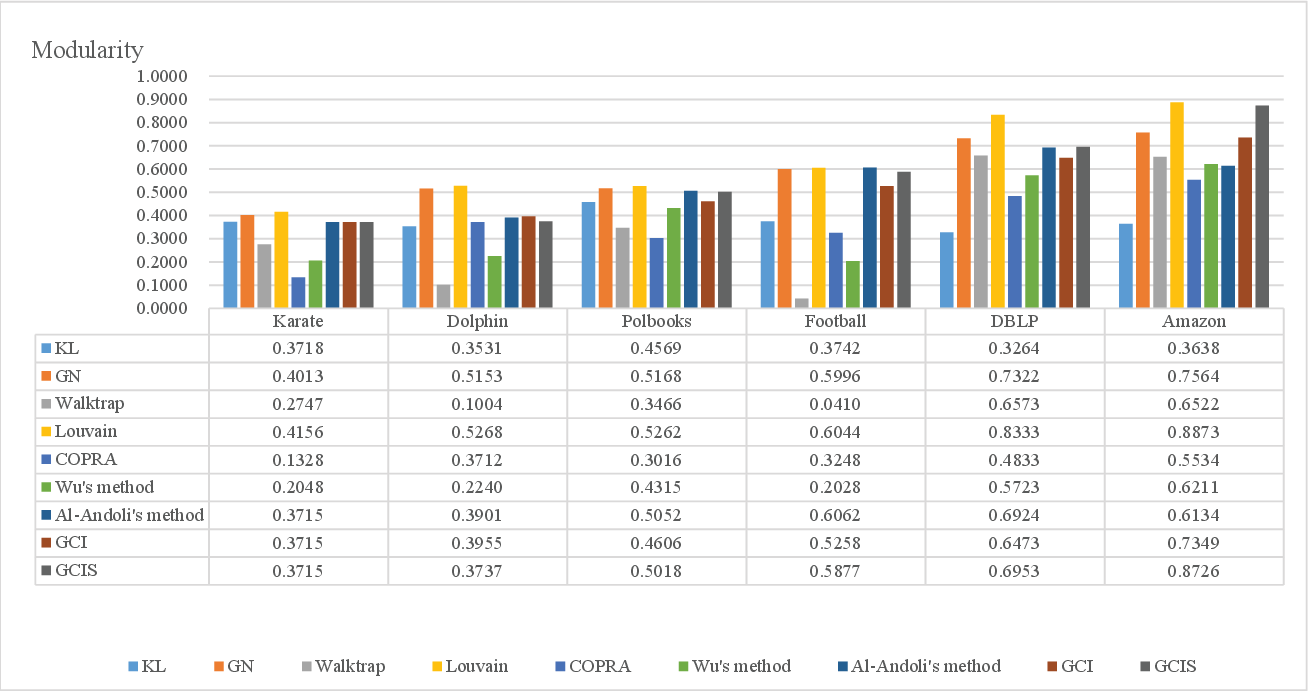}
 \caption{Modularities obtained by different methods in the real-world networks.}
 \label{Modilarity_real}
\end{figure*}

\begin{figure*}
  \centering
  \includegraphics[height=7cm,width=15cm]{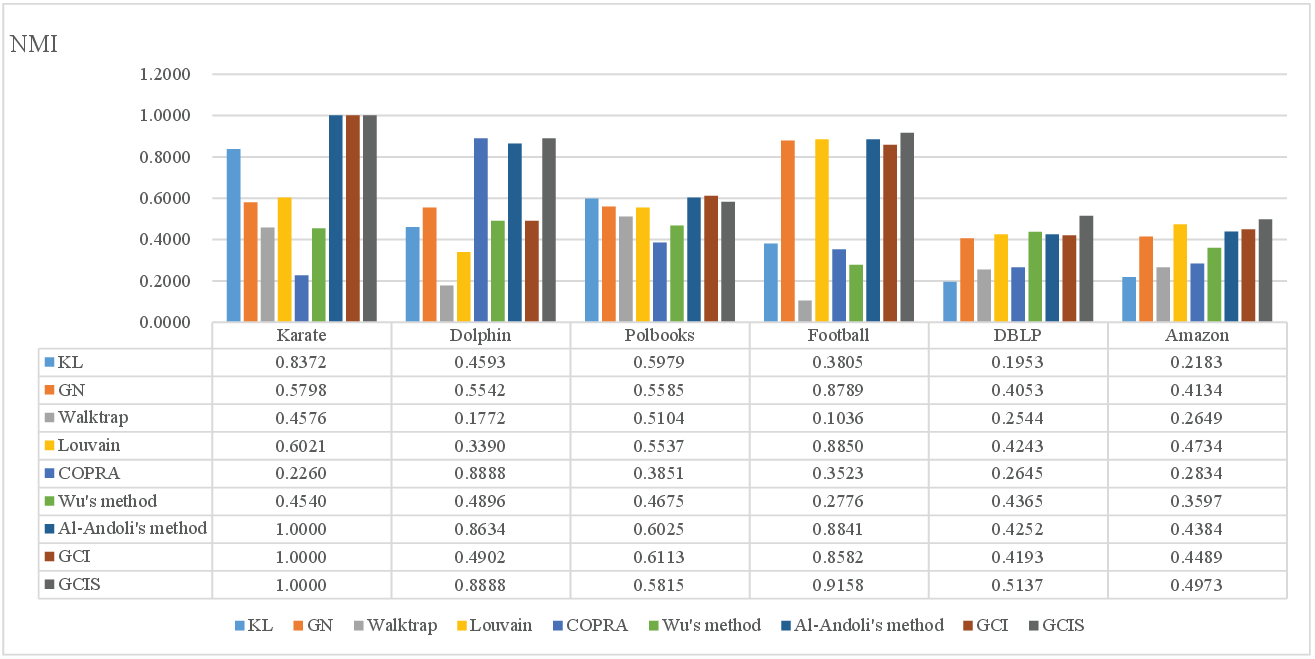}
 \caption{$N\!M\!I$ obtained by different methods in the real-world networks.}
 \label{NMI_real}
\end{figure*}

\begin{figure*}
  \centering
  \includegraphics[height=7cm,width=15cm]{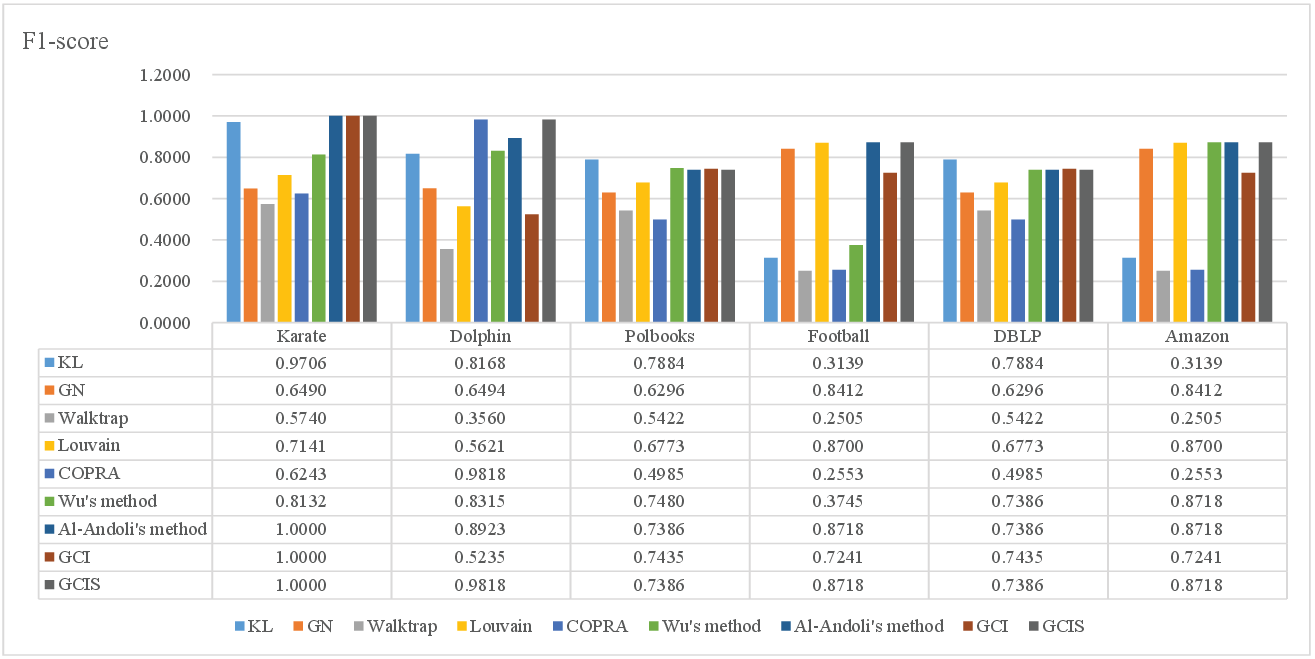}
 \caption{$F1$-score obtained by different methods in the real-world networks.}
 \label{F1_real}
\end{figure*}

The comparative results of our method and the benchmark methods on Modularity, $N\!M\!I$ and $F1$-score for these four networks are shown in Figures \ref{Modilarity_real}, \ref{NMI_real}, \ref{F1_real}.

From Figure \ref{Modilarity_real} it can be seen that, from the perspective of modularity, both Louvain algorithm and GN algorithm have achieved better results. GCIS also achieved high modularity, but it is still slightly lower than the previous three methods. Due to the relatively small number of real communities in these networks, KL has also achieved high modularity. Relatively speaking, the results of Walktrap and COPRA are relatively poor.

From Figures \ref{NMI_real} and \ref{F1_real} it can be seen that, for all networks, our method almost achieve the highest NMI and F1 score values. Especially for karate, GCIS, and GCI, has obtained their true network partitioning. This network has only two communities, but many algorithms partition it into several communities except Al-Andoli's method. 

For network Karate, GCIS, and GCI all obtaiin its groud-truth communities. This network has only two communities, but many algorithms divide it into several smaller communities. 

For Delhpin, compared to the ground-truth communities, the communities obtained by GCIS have only one node with different community affiliation, i.e. node 39. From Figure \ref{Communities_real} (b), it can be seen that node 39 has two edges connecting to nodes 36 and 57 in the two communities, respectively. Therefore, it is also possible for us to divide it into another community.

The advantages of GCIS over GCI are still very obvious, so the random sampling method is also effective in practical networks.

The above experiments fully demonstrate that GCIS has significant advantages compared to some classic community detection algorithms.

\section{Conclusions and future work}

Community structure is an important feature of complex networks. Mining communities can help analyze the structural features and functions of complex networks, and therefore has important theoretical and practical significance. Although various community partitioning methods have been proposed, most of them often rely on the selection of seed nodes.

In view of this, this paper proposes a community detection algorithm based on graph convolution iterative algorithm (GCIS). This algorithm is an iterative algorithm aimed at finding the optimal community partitioning result for the network. The evaluation indicator for community division results is modularity. We have verified the superiority of our method through a large number of experiments.

This paper studies the problem of partitioning non-overlapping communities detection in complex networks. In fact, we can also extend our method to the detection of overlapping communities, which will be the work we need to do in the future.

\section*{CRediT authorship contribution statement}
 \textbf{Jiaqi Yao}: Conceptualization, Investigation, Formal analysis, Methodology, Writing $-$ original draft \& editing.  \textbf{Lewis Mitchell}: Conceptualization, Funding acquisition, Formal analysis, Writing $-$ review, Supervision, Critical revision.

\section*{Declaration of competing interest}
The authors declare that they have no known competing financial interests or personal relationships that could have appeared to influence the work reported in this paper.

\section*{Data availability}
Data will be made available on request.


\bibliography{reference}

\end{multicols}
\end{document}